\documentclass[preprint,12pt]{elsarticle}

\usepackage{amssymb}
\usepackage{amsthm}

\usepackage{soul}

\usepackage{ulem}

\usepackage{newunicodechar}
\newunicodechar{ϵ}{\ensuremath{\epsilon}}

\usepackage{hyperref}
\usepackage{subcaption}
\usepackage{graphicx}
\usepackage{amsmath} 
\usepackage{multicol}
\usepackage{tikz}

\usepackage{mathtools}
\usepackage{comment}
\usepackage{upgreek}
\usepackage{mathrsfs}
\usepackage[flushleft]{threeparttable}
\newcommand{\dd}{\mathrm{d}}

\usepackage{xcolor}

\usepackage[a4paper, total={5.8in, 8.2in}]{geometry}

\journal{ }

\begin{document}

\begin{frontmatter}

\title{Preferential and differential diffusion in RANS simulation of lean hydrogen flames with tabulated chemistry}


\author[label1]{Alex M. Garcia}
\author[label2]{Emiliano M. Fortes}
\author[label2]{Eduardo J. Pérez-Sánchez}
\author[label2]{Daniel Mira}
\author[label3]{Marco Vivenzo}
\author[label3]{Michael Gauding}
\author[label3]{Heinz Pitsch}
\author[label1]{Nico Schmitz}
\author[label1]{Christian Wuppermann}

\address[label1]{Department for Industrial Furnaces and Heat Engineering, RWTH Aachen University, Kopernikusstr. 10, 52074 Aachen, Germany}
\address[label2]{Barcelona Supercomputing Center (BSC), Plaza Eusebi Güell 1-3, 08034, Barcelona, Spain}
\address[label3]{Institute for Combustion Technology, RWTH Aachen University, Templergraben 64, 52056 Aachen, Germany}

\begin{abstract}

Lean hydrogen flames are prone to thermo-diffusive instabilities due to preferential and differential diffusion effects, posing significant challenges for their modeling in computational fluid dynamics simulations. This work extends a tabulated-chemistry (TC) model that includes preferential and differential diffusion effects to a Reynolds-averaged Navier-Stokes (RANS) framework and assesses its performance for a lean premixed $\mathrm{H_2}$-air slot burner at two Reynolds numbers ($\mathrm{Re}=5500$ and $11000$) using direct numerical simulation (DNS) as a reference. The approach is based on transport equations for the progress variable and mixture fraction derived from the species mass transport equations considering mixture-averaged diffusion and Soret effect, and incorporates turbulence--chemistry interaction via a presumed probability density function (PDF) approach. RANS simulations including preferential-differential diffusion are able to correctly reproduce the DNS flame length, heat-release distribution, and the characteristic equivalence-ratio and super-adiabatic temperature branches of the slot flame. Comparisons with (i) a unity-Lewis-number variant and (ii) a model including thermo-diffusive effects only in the flamelet table show the impact of preferential and differential diffusion on the TC model at both the thermochemical and transport levels. Finally, the impact of the turbulence closures for turbulent diffusion, scalar dissipation rate, and Reynolds stresses is assessed. The results presented in this paper demonstrate the capability of the model to include preferential and differential diffusion effects in cost-effective RANS simulations of lean hydrogen flames.

\end{abstract}


\end{frontmatter}


\section{Introduction}
\label{S:1}

Hydrogen combustion has been a major subject of study over the past decade due to its potential as a carbon-free energy solution for hard-to-abate sectors such as power generation, transportation, and industry. However, its practical implementation remains challenging because of several inherent properties, including high flame speed and mass diffusivity, low volumetric energy density, a tendency to produce NO$_x$, and increased susceptibility to combustion instabilities. These characteristics complicate its use in industrial combustion systems and require careful consideration in both design and modeling \cite{SongFranc23}.

One of the key aspects of hydrogen combustion modeling is the role of transport phenomena resulting from hydrogen's high mass diffusivity. In lean hydrogen flames, two critical transport-related effects arise: differential diffusion, caused by the significant disparity between thermal and mass diffusivities (non-unity Lewis number), and preferential diffusion, resulting from the different mass diffusivities of the species involved in the combustion process \cite{Matal07, Pitsc24}. While these effects are not unique to hydrogen, their impact is particularly pronounced in hydrogen combustion. They significantly influence flame propagation and promote thermo-diffusive instabilities, especially under very lean conditions, as targeted in certain applications aimed at reducing NO$_\mathrm{x}$.

Tabulated chemistry (TC) models are a family of approaches based on the assumption that a few control variables, e.g., a mixture fraction and a reaction progress variable, can fully describe the thermochemical state of the reacting mixture. In other words, it is assumed that any flame state lies on a low-dimensional manifold that can be precomputed and stored in a table for later use. The most notable examples of tabulated chemistry methods include the Flamelet-Generated Manifold (FGM)~\cite{vanODonin16}, the Flame Prolongation of Intrinsic low-dimensional manifolds (FPI)~\cite{GicquDarab00}, and the Flamelet Progress Variable (FPV)~\cite{KnudsPitsc09} approaches. Over the years, significant progress has been made in developing TC models that incorporate preferential and differential diffusion effects~\cite{SwartBasti10, DoninBasti15, RegelKnuds13, MukunEfimo21, BottlChen22, BottlLulic23, PerezForte25}. A more challenging task is the extension of such models to turbulent combustion, as thermo-diffusive instabilities synergistically interact with turbulence~\cite{Pitsc24}. 

A pioneering contribution to the inclusion of differential diffusion effects in TC models was made by Vreman et al. \cite{VremavanO09}, who determined effective Lewis numbers for the control variables by analyzing the spatial structure of the flame. Building on this foundation, subsequent studies introduced additional transport terms into the governing equations of the control variables to explicitly capture fluxes arising from the complete set of control variables \cite{SwartBasti10,DoninBasti15}. 

Regele et al. \cite{RegelKnuds13} proposed a new approach based on a modified mixture fraction equation which includes a progress-variable-dependent source term assuming one-step chemistry and unity Lewis number for all the species except the fuel. This approach was successfully applied to capture the dynamics of spherical hydrogen and propane flames, showing good agreement with reference data. Schlup et al. \cite{SchluBlanq19} later extended the approach by including mixture-averaged diffusion and the Soret effect, further enhancing its predictive capabilities.

De Swart et al. \cite{SwartBasti10} and Donini et al. \cite{DoninBasti15} derived modified transport equations for the control variables that account for differential and preferential diffusion effects by applying the chain rule to the species fluxes and linearly combining the species mass fractions. These equations included new cross terms representing mutual interactions among the control variables, which were ultimately consolidated into a single term dependent only on the progress variable flux, in accordance with the flamelet hypothesis.

Böttler et al. \cite{BottlChen22} introduced an alternative strategy in which a selected subset of species mass fractions is transported and subsequently used to reconstruct the control variables for manifold access. This approach was applied to predict the dispersion relation of expanding spherical flames \cite{BottlLulic23} and was compared against the method of Wen et al. \cite{WenZirwe22,WenZirwe22b}, which explicitly accounted for curvature and strain rate effects. Both methods yielded accurate predictions at low wavenumbers, though discrepancies arose at intermediate and high wavenumbers. Mukundakumar et al. \cite{MukunEfimo21} proposed a method based on commuting operators under the assumption of constant Lewis numbers to derive new coefficients for the control variables. This formulation improved the prediction of slit flame behavior compared to the results of Donini et al. \cite{DoninBasti15}. Finally, the approach of Donini et al. \cite{DoninBasti15} was extended by Pérez-Sánchez et al. \cite{PerezForte25} by incorporating a mixture-averaged transport model when deriving the transport equations for the control variables and building the flamelet database, leading to a robust and comprehensive tabulated chemistry model that includes preferential and differential diffusion effects. This model was used in a comprehensive assessment of the flamelet hypothesis and its implications for canonical hydrogen-based configurations \cite{PerezForte25,FortePerez25b}, highlighting the method's capability to capture thermo-diffusive effects.

In the case of turbulent flames, the flamelet hypothesis assumes that a turbulent flame consists of an ensemble of one-dimensional laminar flames, leading to a reduction in the dimensionality of the problem and allowing the application of manifolds constructed from one-dimensional laminar flames. The unresolved turbulence\--chemistry interaction is typically incorporated by using a presumed Probability Density Function (PDF) approach. The tabulated chemistry models, coupled with presumed PDFs, have been widely used in numerical simulations of turbulent combustion systems within both Reynolds-Averaged Navier–Stokes (RANS) and Large Eddy Simulation (LES) frameworks. However, despite the substantial advancements, the inclusion of thermo-diffusive effects in tabulated chemistry models for turbulent premixed combustion remains an active area of research \cite{DoninBasti17, ZhangKarac21, NicolDress22, KaiTokuo23, BallaSomer25, FortePerez25, BergeAttil25}.

A common approach to assess the ability of the TC models to accurately capture thermo-diffusive effects in turbulent flames is to compare their predictions against direct numerical simulations (DNS) incorporating detailed transport and chemistry. For example, Ferrante et al. \cite{FerraEitel24} performed LES of the lean hydrogen\-–air premixed slot burner investigated by Berger et al. \cite{BergeAttil22} using the models of Regele et al. \cite{RegelKnuds13} and Mukundakumar et al. \cite{MukunEfimo21}, both of which overpredicted flame lengths relative to the DNS data. More recent studies by Berger et al. \cite{BergeAttil25} and Fortes et al. \cite{FortePerez25} extended the models of Schlup et al. \cite{SchluBlanq19} and Pérez-Sánchez et al. \cite{PerezForte25}, respectively, and applied them to LES of the same slot burner. These models showed improved predictions of the flame length, as well as the temperature and equivalence ratio fields.


Regarding RANS turbulence modeling, limited research has explored the use of tabulated chemistry combustion models that account for preferential and differential diffusion \cite{LiuShao21, AlmutDines23}. Nevertheless, RANS remains a valuable tool for engineering applications due to its low computational cost, making it essential to evaluate the ability of such models to capture key features of lean hydrogen turbulent flames. In this context, the present work extends the tabulated chemistry model of Pérez-Sánchez et al. \cite{PerezForte25} to RANS simulations and evaluates its performance against DNS data for a lean hydrogen\-–air premixed slot burner at two different Reynolds numbers. The objective of the present work is not only to evaluate the accuracy of the RANS simulations using the TC model but also to assess how the inclusion of preferential and differential diffusion affects the different aspects of the TC model. To this end, additional simulations are conducted using the TC model with a unity Lewis number and with preferential and differential diffusion effects applied only at the thermochemical level. Moreover, freely propagating one-dimensional laminar flames are also computed and analyzed to further clarify the behavior of the model in the turbulent flames. Finally, the influence of the turbulence, the turbulent diffusion, and the turbulent scalar dissipation rate modeling on the RANS simulation of the slot flame is examined.

\section{Model description} 
\label{S:2}

Modeling thermo-diffusive instability effects requires an accurate representation of thermal and mass diffusive transport. To this end, the present work employs a tabulated chemistry method derived from the species mass conservation equation using a mixture-averaged transport model, incorporating thermal diffusion (Soret effect) and a velocity correction to ensure mass conservation \cite{PerezForte25, FortePerez25b}. 

The mass diffusion velocity of each species $k$ with a non-zero mass fraction, expressed according to the mixture-averaged model and including the Soret effect, is given by:

\begin{equation}\label{eq:diff_eq_MA}
\mathbf{V}_k = - \frac{D_k}{Y_k} \frac{W_k}{W}\nabla X_k - \frac{D_k^T}{Y_k}\frac{\nabla T}{\rho T}, \quad \quad D_k = \frac{1-Y_k}{\sum_{\substack{j=1 \\ j \neq k}}^{N_s} X_j / \mathcal{D}_{jk}}, \quad \quad k=1,\ldots,N_s.
\end{equation}

\noindent where $D_k$ is the diffusion coefficient of the k-th species while $\mathcal{D}_{jk}$ are the binary diffusion coefficients between species $j$ and $k$, $Y_k$ is the mass fraction, $W_k$ is the molecular weight, $X_k$ is the molar fraction, $D_k^T$ is the thermal diffusion coefficient for species $k$, $W$ is the mean molecular weight of the mixture, and $N_s$ is the number of species \cite{PoinsVeyna05}. 
The mass diffusion flux for each species after introducing the velocity correction is given by:


\begin{align}
    \mathbf{j}_k = \rho Y_k (\mathbf{V}_k+\mathbf{V}_c) =  & \rho \left[ -D_k \frac{W_k}{W}\nabla X_k
    +\frac{Y_k}{W}\sum_{j=1}^{N_s}D_j\,W_j\,\nabla X_j - \frac{D_k^T}{\rho \, T}\nabla T + \right. \notag
    \\& \left. \frac{Y_k}{\rho\,T}\nabla T\sum_{j=1}^{N_s}D_j^T \right],  \quad \quad k=1, \ldots,N_s.
    \label{eq:diff_eq_MA_total}
\end{align}


The flamelet hypothesis assumes that any thermochemical state within a flame lies on a manifold defined as a function of a reduced set of control variables $\{C_i\}_{i=1}^{N_c}$. This manifold is constructed by assembling a database of representative flamelets, typically one-dimensional laminar flames, covering the range of mixture compositions and reaction states. If the diffusive fluxes are expressed as functions of the manifold control variable fluxes, new transport equations can be derived for the control variables. To this end, the gradients can be reformulated by applying the chain rule as $\nabla \psi = \sum_{i=1}^{N_c} \nabla C_i \frac{\partial \psi}{\partial C_i}$ where $\psi$ is any dependent variable of the manifold. These new transport equations entirely retain the preferential and differential diffusion effects, since no assumption is made on the form of the species fluxes, and are valid as long as the variables remain on a manifold. The reader is referred to \cite{PerezForte25} for a complete derivation and description of the modeling approach.

In the current work, the control variables are a reaction progress variable and a mixing variable, as no heat losses or gains are considered. On the one hand, the transition from unburned to burned conditions is described by the progress variable $Y_c$, defined by a linear combination of species mass fractions: $Y_c = \sum_{k=1}^{N_s} a_k Y_k$, where $a_k$ are weighting constants chosen to ensure a monotonic variation of $Y_c$ along the flame spatial coordinate. On the other hand, the local mixture composition is described by the Bilger's mixture fraction $Z$:

\begin{equation}\label{eq:Zbilger}
Z=\frac{\beta-\beta_{\mathrm{ox}}}{\beta_{\mathrm{fuel}}-\beta_{\mathrm{ox}}}, \qquad \beta=2 \frac{Z_C}{W_C}+\frac{1}{2} \frac{Z_H}{W_H}-\frac{Z_O}{W_O},
\end{equation}

\noindent where $Z_{C}$, $Z_{H}$ and $Z_{O}$ denote the elemental mass fractions of carbon, hydrogen, and oxygen, while $W_{C}$, $W_{H}$ and $W_{O}$ represent their corresponding atomic weights. The transport equations for the progress variable and mixture fraction are obtained by linear combination of the species mass fraction transport equations according to each definition, leading to:

\begin{equation}
    \frac{\partial (\rho Y_c)}{\partial t} +  \nabla \cdot \left(\rho \mathbf{u} ~Y_c\right) =
    \nabla \cdot \left( \rho \Gamma_{Y_c,Y_c} \nabla Y_c + \rho \Gamma_{Y_c,Z} \nabla Z \right)
    + \rho \dot{\omega}_{c},
    \label{Eq_Yc} 
\end{equation}

\begin{equation}
    \frac{\partial (\rho Z)}{\partial t} + \nabla \cdot \left(\rho \mathbf{u} ~Z\right) =
    \nabla \cdot \left( \rho \Gamma_{Z,Y_c} \nabla Y_c + \rho \Gamma_{Z,Z} \nabla Z \right),
    \label{Eq_Z} 
\end{equation}

\noindent where coefficients $\Gamma_{Y_c,Y_c}$, $\Gamma_{Y_c,Z}$, $\Gamma_{Z,Y_c}$ and $\Gamma_{Z,Z}$ are the molecular diffusion coefficients that relate gradients of $Y_c$ and $Z$ to the diffusive fluxes of $Y_c$ and $Z$. They are computed from the flamelet database by applying the chain rule to the detailed species diffusion fluxes and linearly combining them according to the definitions of $Y_c$ and $Z$. These diffusion coefficients are stored in the lookup table together with all the thermochemical variables of the flamelet database \cite{PerezForte25}. Note that preferential\--differential diffusion effects lead to two types of molecular diffusion terms: normal diffusion, proportional to the control variable’s own gradient, and cross-diffusion, proportional to the gradient of the other control variables. These cross-diffusion terms act as source/sink terms. This is particularly relevant in the case of the mixture fraction, which would otherwise behave as a passive scalar if the term $\nabla \cdot (\rho \Gamma_{Z,Y_c} \nabla Y_c)$ was neglected.

For storing data in the manifold, it is advantageous to use variables that range between 0 and 1, so that the flame states are confined within a region defined by a square (or, more generally, a hypercube). To this end, the normalized progress variable is defined as:

\begin{equation}\label{eq:def_c}
    c = \frac{Y_c-Y_{c,u}(Z)}{Y_{c,b}(Z)-Y_{c,u}(Z)}.
\end{equation}

\noindent where the subscripts $u$ and $b$ denote the unburned and burned sides of the flamelet, respectively.

Note that, since no control variable is included for heat transfer, the enthalpy is fully determined by the progress variable and the mixture fraction, and no heat loss or gain is considered.

\subsection{Turbulence\--chemistry interaction} 

In the RANS simulations, the transport equations are solved in Favre-averaged (density-weighted time-averaged) form. This leads to the following transport equations for the mean progress variable $\widetilde{Y}_c$ and the mean mixture fraction $\widetilde{Z}$:

\begin{equation}
     \frac{\partial (\overline{\rho} \, \widetilde{Y}_c)}{\partial t} 
     + \nabla \cdot \left(\overline{\rho} \, \widetilde{\mathbf{u}} \, \widetilde{Y}_c\right) 
     = \nabla \cdot \left[ \overline{\rho} \left(\widetilde{\Gamma}_{Y_c,Y_c} + \frac{\nu_t}{Sc_{t,c}} \right)\,\nabla \widetilde{Y}_c \right]
     + \nabla \cdot \biggl(\overline{\rho} \, \widetilde{\Gamma}_{Y_c,Z} \,\nabla \widetilde{Z}\biggr)  
     + \overline{\rho} \, \widetilde{\dot{\omega}}_c,
    \label{Eq_turb_Yc} 
\end{equation}

\begin{equation}
    \frac{\partial (\overline{\rho} \, \widetilde{Z})}{\partial t} 
    + \nabla \cdot \left(\overline{\rho} \, \widetilde{\mathbf{u}} \, \widetilde{Z}\right) 
    = \nabla \cdot \left[ \overline{\rho} \left( \, \widetilde{\Gamma}_{Z,Z} + \frac{\nu_t}{Sc_{t,z}} \right)\,\nabla \widetilde{Z} \right] 
    + \nabla \cdot \biggl( \overline{\rho} \, \widetilde{\Gamma}_{Z,Y_c} \,\nabla \widetilde{Y}_c\biggr) 
    \label{Eq_turb_Z} 
\end{equation}

\noindent where Favre averages are denoted by $\widetilde{\cdot}$ and Reynolds averages by $\overline{\cdot}$. The effect of turbulence on the convective fluxes of the control variables is modeled using the gradient diffusion hypothesis, which treats unresolved turbulent transport as turbulent diffusion, analogously to molecular diffusion. Accordingly, the turbulent diffusion of $\widetilde{Y_c}$ and $\widetilde{Z}$ is defined as proportional to their gradients, with an effective diffusivity given by the turbulent viscosity $\nu_t$ and a turbulent Schmidt number $Sc_{t}$.

Turbulence\--Chemistry Interaction (TCI) is accounted for by integrating the laminar manifold over a family of presumed $\beta$-shaped PDFs, $\mathcal{P}$, creating a ``turbulent manifold" that incorporates the effects of turbulence on the flamelets. These PDFs are parameterized by the first and second moments of the control variables: the mean normalized progress variable $\widetilde{c}$ and mean mixture fraction $\widetilde{Z}$, and their respective variances, $c_v$ and $Z_v$. For simplicity, statistical independence between the mixture fraction and the progress variable is assumed. However, it should be noted that this assumption may not be entirely valid in the presence of thermo-diffusive instabilities and should be examined in future work. Therefore, in the current work, any thermochemical variable $\Psi$ is given by:

\begin{equation}
\begin{split}
    \widetilde{\Psi}(\widetilde{c}, c_v,\widetilde{Z}, Z_v)  =\int_0^1 \int_0^1 \Psi (c,Z) \mathcal{P}_c(c;\widetilde{c}, c_v) \, \mathcal{P}_Z(Z; \widetilde{Z}, Z_v) \, \dd c\, \dd Z.
\end{split}
\end{equation}

\noindent 



The transport equations for the variance of the progress variable $Y_{c,v}$ and variance of the mixture fraction $Z_v$ are derived from the instantaneous transport equations for $Y_c$ and $Z$ (Equations.~\ref{Eq_Yc} and \ref{Eq_Z}) by applying Favre decomposition and Favre averaging, and by manipulating the resulting relations to obtain transport equations for the second-order moments of the
fluctuations, leading to:

\begin{align}
    \frac{\partial (\overline{\rho} \, Y_{c,v})}{\partial t} 
    + \nabla \cdot \left(\overline{\rho} \, \widetilde{\mathbf{u}} \, Y_{c,v}\right) 
    =& \nabla \cdot \left[ \overline{\rho} \left(\widetilde{\Gamma}_{Y_c,Y_c} + \frac{\nu_t}{Sc_{t,c}} \right) \nabla Y_{c,v} \right]
    + 2 \overline{\rho} \, \frac{\nu_t}{Sc_{t,c}} \nabla \widetilde{Y}_c \cdot \nabla \widetilde{Y}_c  \notag
    \\& - 2 \overline{\rho} \, C_{d,c} \frac{\epsilon}{\kappa} Y_{c,v} 
    + 2 \overline{\rho} \left( \widetilde{Y_c \dot{\omega}_c} - \widetilde{Y}_c \widetilde{\dot{\omega}}_c \right),
    \label{Eq_turb_varYc} 
\end{align}

\begin{align}
    \frac{\partial (\overline{\rho} \, Z_{v})}{\partial t} 
    + \nabla \cdot \left(\overline{\rho} \, \widetilde{\mathbf{u}} \, Z_{v}\right) 
    =& \nabla \cdot \left[ \overline{\rho}\left( \widetilde{\Gamma}_{Z,Z} + \frac{\nu_t}{Sc_{t,z}} \right) \nabla Z_{v} \right]
    + 2 \overline{\rho} \, \frac{\nu_t}{Sc_{t,z}} \nabla \widetilde{Z} \cdot \nabla \widetilde{Z} \notag 
    \\& - 2 \overline{\rho} \, C_{d,z} \frac{\epsilon}{\kappa} Z_{v} .
    \label{Eq_turb_varZ} 
\end{align}

The scalar dissipation rates of $Y_{c,v}$ and $Z_v$ are respectively modeled as $2 \overline{\rho} \,C_{d,c} \, (\epsilon/\kappa) \, Y_{c,v}$ and $2 \overline{\rho} \,C_{d,z} \, (\epsilon/\kappa) \, Z_v$ \cite{PoinsVeyna05}, where $C_{d,c}$ and $C_{d,z}$ are model constants that relate the scalar mixing and turbulent time scales. The turbulent kinetic energy $\kappa$ and its dissipation rate $\epsilon$ are computed using a RANS turbulence model. Both the Favre-averaged chemical source term $\widetilde{\dot{\omega}}_c$ and the Favre-averaged reaction--progress moment $\widetilde{Y_c \dot{\omega}_c}$ are obtained alongside all thermochemical variables by the PDF integration of the laminar manifold.

Similar to the laminar case, the normalized progress variable $\widetilde{c}$ and its variance $c_v$ are used to facilitate manifold tabulation, and they are computed as:

\begin{equation}\label{eq:c_turb}
    \widetilde{c} = \frac{\widetilde{Y}_c - \widetilde{Y}_{c,u}(\widetilde{Z},Z_v)}{\widetilde{Y}_{c,b}(\widetilde{Z},Z_v)-\widetilde{Y}_{c,u}(\widetilde{Z},Z_v)},
\end{equation}

\begin{equation}\label{eq:c_v_turb}
    c_v = \frac{Y_{c,v} + \widetilde{Y}_c^2 - \widetilde{Y_{c,u}^2} - 2(\widetilde{Y_{c,u} Y_{c,b}}-\widetilde{Y_{c,u}^2})\,\widetilde{c}}{\widetilde{Y_{c,b}^2} - 2\widetilde{Y_{c,u} Y_{c,b}}+\widetilde{Y_{c,u}^2}} - \widetilde{c}^2,
\end{equation}

\noindent where the values of the progress variable on the unburned and burned sides of the flamelet, $\widetilde{Y}_{c,u}$ and $\widetilde{Y}_{c,b}$, are obtained by PDF integration and tabulated as functions of $\widetilde{Z}$ and $Z_v$. The variances $c_v$ and $Z_v$ are normalized to range between 0 and 1, as follows:

\begin{equation}\label{eq:c_Sc}
    S_C = \frac{c_v}{\widetilde{c} \left(1- \widetilde{c}\right)}, \quad \mathrm{and } \quad S_Z = \frac{Z_v}{\widetilde{Z} (1- \widetilde{Z})}.
\end{equation}

Finally, the turbulent manifold, which contains the averaged values $\widetilde{\Psi}$ corresponding to the different quadruples $(\widetilde{c}, S_C, \widetilde{Z}, S_Z)$, is accessed as a lookup table during the turbulent flame simulation.

\section{Reference case and DNS numerical setup}
\label{S:3}
The reference case used to evaluate the TC model with preferential and differential diffusion is the slot burner configuration studied by Berger et al. \cite{BergeAttil22,BergeAttil24}. A fully premixed hydrogen-air mixture with an equivalence ratio of 0.4 and an inlet temperature of $T_u = 298$ K is injected through two parallel plates separated by a distance $H$. The main injection is surrounded by a hot coflow consisting of burned gases from the combustion of the same hydrogen-air mixture, i.e., with the same equivalence ratio and an adiabatic flame temperature of $T_b = 1416$ K. The coflow is injected at a velocity of 3.6 m/s, while the main jet has a bulk velocity of 24 m/s. The coflow and main jet are separated by walls of thickness $H/20$, at a fixed temperature of $T_{\text{wall}} = 298$ K. Two conditions are evaluated, corresponding to Reynolds numbers of 5500 and 11000, defined as Re $= v_{bulk} \,H /\nu_u$, where $v_{bulk}$ is the bulk velocity and $\nu_u$ the kinematic viscosity of the unburned mixture. These conditions are associated with wall distances of $H = 4$ mm and $H = 8$ mm, respectively.

DNS of the slot burner configuration is performed using the models and numerical methods described in detail by Berger et al. \cite{BergeAttil22,BergeAttil24}. A multi-species ideal gas is computed by solving the low-Mach Navier–Stokes equations. Chemical reactions are modeled using the detailed mechanism of Burke et al. \cite{BurkeChaos12}, which includes 9 species and 46 reactions. Thermal conductivity and viscosity are computed from kinetic gas theory using mixture-averaged rules. Species diffusion coefficients are evaluated using the constant non-unity Lewis number approach, including thermal diffusion.


The three-dimensional (3D) domain extends in the cross-wise, streamwise, and spanwise directions (corresponding to $x$, $y$, and $z$, respectively) by $13H$, $15.5H$, and $4.6H$ for the case of $\text{Re} = 11000$. For $\text{Re} = 5500$, the streamwise domain extends to $18H$ to accommodate the longer flame length. The domain is discretized with a resolution of 70 $\mu$m to ensure that the Kolmogorov scale is well resolved at all times and locations, while maintaining at least 10 grid points across the thermal flame thickness of the reference 1D unstretched laminar flame. Periodic boundary conditions are imposed in the spanwise direction, while symmetry boundary conditions are used in the cross-wise direction. Additionally, an auxiliary channel flow simulation is employed to generate the turbulent inlet conditions at the jet inlet.


\begin{figure}[h!]
\centering
\includegraphics[trim = 0 190 550 0  ,clip,width=0.55\textwidth]{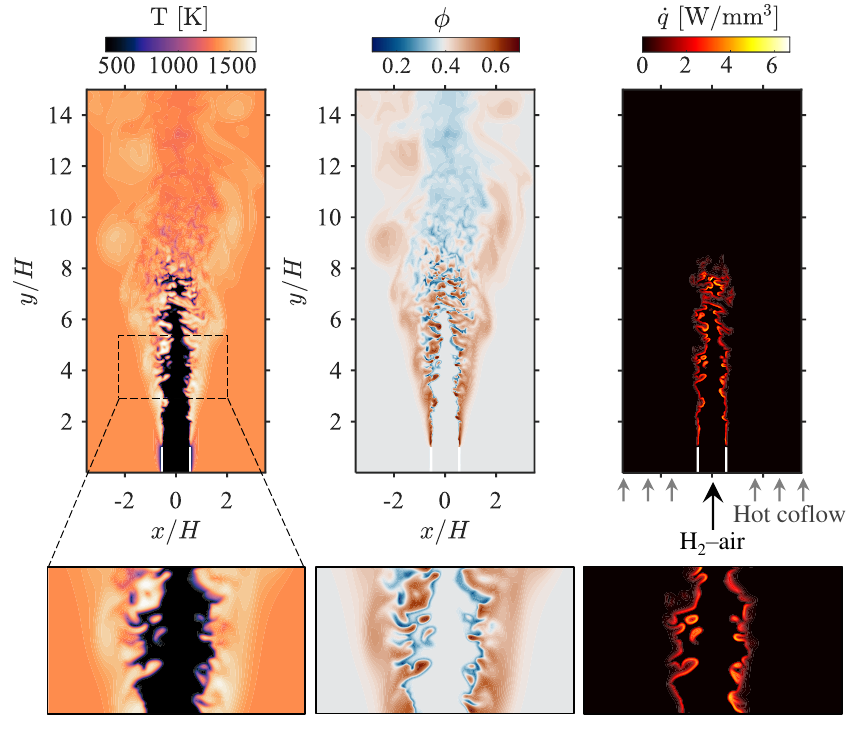}
\caption{Instantaneous fields of temperature, equivalence ratio, and heat release rate from the DNS of the hydrogen-air slot flame with Re = 11000.}
\label{Fig_DNS_flame}
\end{figure}

The temperature, equivalence ratio, and heat release rate fields for the turbulent slot flame at $\text{Re} = 11000$ are shown in Figure~\ref{Fig_DNS_flame}. Regions of super-adiabatic temperature (i.e., temperatures exceeding the adiabatic flame temperature at the global equivalence ratio of the premixed mixture) arise due to preferential-differential diffusion effects in the lean hydrogen-air flame. The preferential diffusion results in local variations in the equivalence ratio and, consequently, the adiabatic flame temperature.

The equivalence ratio decreases ahead of the flame front because $\text{H}_2$ diffuses faster than $\text{O}_2$. Additionally, flame curvature and strain locally focus (when positive) or defocus (when negative) the diffusion of $\text{H}_2$, generating local richer or leaner mixtures, respectively \cite{ChungLaw88, Law89, AltanFrouz12}. This enhances the heat release rate in regions of positive curvature, causing intense burning, while reducing it in regions of negative curvature \cite{ChenIm00}. Moreover, differential diffusion of heat and mass ahead of the flame similarly affects enthalpy, further varying the heat release rate along the turbulent flame \cite{Matal07, Pitsc24}.

\section{RANS numerical setup}
\label{S:4}

Simulations of the turbulent premixed hydrogen-air slot flames are performed using the TC model with preferential and differential diffusion by solving Equations~\eqref{Eq_turb_Yc}-\eqref{Eq_turb_varZ} alongside the Reynolds-averaged Navier–Stokes (RANS) transport equations for continuity and momentum. The transport equations are solved using the pressure-based solver in ANSYS Fluent v2023-R1, employing a pseudo-time-stepping coupled scheme for pressure-velocity coupling and the PRESTO! (PREssure STaggering Option) scheme for pressure interpolation \cite{Patan18}. A second-order upwind spatial discretization scheme is applied to all transport equations.

The density and viscosity of the gas mixture, as well as all diffusion coefficients, are extracted from the lookup tables based on the control variables $Y_c$, $Z$, $Y_{c,v}$, and $Z_v$. The transport equations for these control variables are implemented in the solver as user-defined scalars. The RANS version of the tabulated chemistry model introduces two constants for each pair of control variables $(\widetilde{Y}_c, Y_{c,v})$ and $(\widetilde{Z}, Z_v)$: the turbulent Schmidt number ($Sc_{t,c}$ and $Sc_{t,z}$) and the constant for the scalar dissipation rate model ($C_{d,c}$ and $C_{d,z}$). The impact of these model constants is assessed in Sections~\ref{S:TurbDiff} and~\ref{S:TurbSDR}. The values used for the simulation are $Sc_{t,c}$ = $Sc_{t,z}$ = 0.5, and $C_{d,c}$ = $C_{d,z}$ = 3.0. The flamelets for the chemistry tabulation are computed from one-dimensional freely propagating premixed flames, using the same chemical mechanism as in the DNS. The progress variable is defined based on the H$_2$O mass fraction, such that $Y_c = Y_{\mathrm{H_2O}}$.

The Reynolds stress tensor in the RANS transport equations is computed using the Boussinesq hypothesis with the $\kappa$–$\omega$ turbulence model proposed by Wilcox \cite{Wilco88}. The kinematic turbulent viscosity $\nu_t$ is computed as the ratio of the turbulence kinetic energy ($\kappa$) to its specific dissipation rate ($\omega$), i.e., $\nu_t = \kappa/\omega$, and transport equations are solved for $\kappa$ and $\omega$, as follows:

\begin{equation}
    \frac{\partial (\overline{\rho}\,\kappa)}{\partial t}
    \;+\;\nabla \cdot (\overline{\rho} \, \widetilde{\mathbf{u}}\,\kappa)
    \;=\;
    \nabla \cdot \left[ \overline{\rho}\, \Bigl( \widetilde{\nu} + \frac{\nu_t}{\sigma_{\kappa}}\Bigr)\,\nabla \kappa \right] + 2\overline{\rho}\,\nu_t\,\mathbf{S}:\mathbf{S} - \overline{\rho}\,\beta^*\,f_{\beta*}\,\kappa\,\omega , 
    \label{Eq_kappa}
\end{equation}

\begin{equation}
    \frac{\partial (\overline{\rho}\,\omega)}{\partial t}
    \;+\;\nabla \cdot (\overline{\rho} \, \widetilde{\mathbf{u}}\,\omega)
    \;=\;
    \nabla \cdot \left[ \overline{\rho}\, \Bigl( \widetilde{\nu} + \frac{\nu_t}{\sigma_{\omega}}\Bigr)\,\nabla \omega \right] + 2\,\alpha\,\overline{\rho}\,\nu_t\,\frac{\omega}{\kappa}\,\mathbf{S}:\mathbf{S} - \overline{\rho}\,\beta\,f_\beta\,\omega^2  ,
    \label{Eq_omega}
\end{equation}

\noindent where $\mathbf{S}$ represents the strain-rate tensor. The values $\alpha$, $\sigma_{\kappa}$, $\sigma_{\omega}$, $\beta^* = f(\beta^*_\infty)$, $\beta = f(\beta_i)$, $f_{\beta*}$, and $f_\beta$ are empirical parameters and correlations, typically determined through calibration. The standard values of $\alpha = 0.52$, $\beta^*_\infty = 0.09$, $\beta_i = 0.072$, and $\sigma_{\kappa} = \sigma_{\omega} = 2$ are used \cite{ANSYS23}. The turbulent kinetic energy dissipation rate is then computed as $ \epsilon = \beta^*\,\kappa\,\omega$. Since the turbulence model plays a crucial role in RANS simulations, its impact on the simulation performance is assessed in Section~\ref{S:TurbModel}. 

A two-dimensional (2D) domain is selected for the RANS simulations due to the periodic boundary conditions in the spanwise direction ($z$-axis) of the slot burner. The domain retains the same streamwise ($y$-axis) and cross-wise ($x$-axis) extents as the DNS domain, except that the inlet jet section is extended upstream by a total distance of $5H$ to ensure full turbulence development within the parallel plates. The domain is discretized using a structured mesh consisting of 44000 quadrilateral elements, with a characteristic cell size of 0.15 mm between the parallel plates and 0.25 mm in the core of the jet injection. Although this resolution is not typical for practical 3D RANS simulations, it is adopted here due to the low computational cost of the 2D approach and the narrow wall thickness of the slot flame. The impact of mesh resolution was evaluated by comparing meshes ranging from 140000 down to 6300 elements, showing very small differences in the results, localized in regions of sharp gradients.

\begin{figure}[h!]
\centering
\includegraphics[trim = 150 323 160 325 ,clip, width=0.5\linewidth]{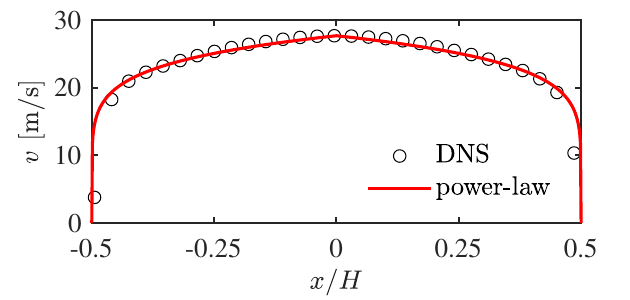}
\caption{Central jet inlet velocity profile.}
\label{Fig_Vin}
\end{figure}

A symmetry condition is applied to both side boundaries in the cross-wise direction, while a zero-gradient condition is imposed at the downstream boundary. The fully premixed hydrogen-air mixture ($Z = 0.0116$ and $Y_c = 0.0$) enters the central jet with a velocity profile using a power-law fitted to the DNS-averaged, as shown in Figure~\ref{Fig_Vin}. The coflow of burned gases ($Z = 0.0116$ and $Y_c = 0.10368$) enters with a constant velocity profile of 3.6 m/s. The turbulence at the central jet inlet is prescribed using constant values for the turbulence intensity and the turbulent length scale, equal to $0.16\,\text{Re}^{-1/8}$ and $0.07(2H)$, respectively \cite{ANSYS23}. Finally, the coflow is specified with zero turbulence intensity.


\section{Results}

To analyze the tabulated chemistry model with preferential and differential diffusion (TC PD) in the RANS simulations, two additional model versions are considered. The first version assumes unity Lewis number (TC Le1), implying that neither preferential nor differential diffusion is included in the flamelet tabulation. Consequently, the diffusion coefficients in the transport equations for the control variables are given by ${\Gamma}_{Y_c,Z} = {\Gamma}_{Z,Y_c} = 0.0$ (null cross-diffusion) and ${\Gamma}_{Y_c,Y_c} = {\Gamma}_{Z,Z} = D_\mathrm{th}$, where $D_\mathrm{th}$ denotes the thermal diffusivity of the mixture. The second version includes preferential and differential diffusion only in the flamelet tabulation (TC PD-F), i.e., at the thermochemical level, but not at the transport level. Thus, the transport equations remain identical to those in the unity Lewis case.

\subsection{Laminar premixed flame}

\begin{figure}[h!]
\centering
\includegraphics[trim = 0 20 730 0 ,clip, width=0.5\linewidth]{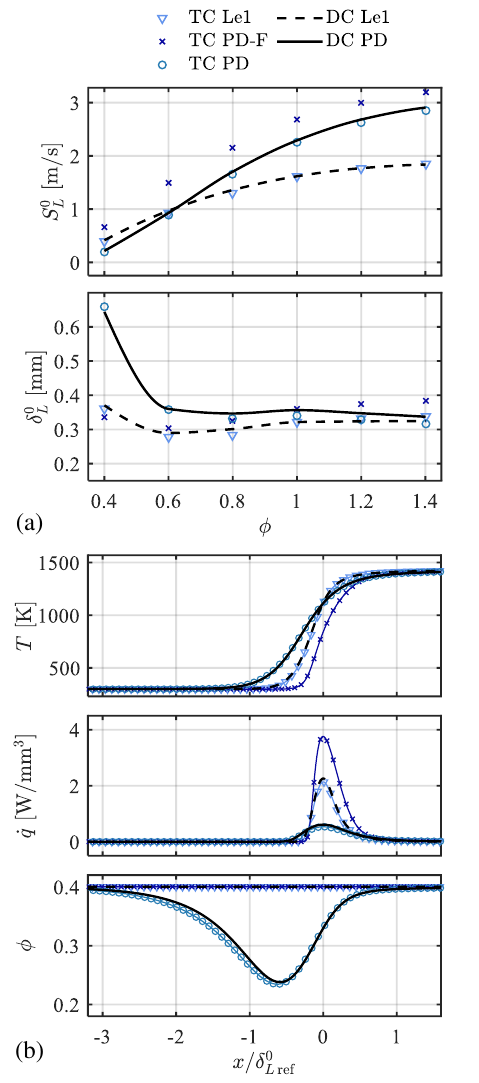}
\caption{(a) Variation of unstretched laminar flame speed and thermal flame thickness with equivalence ratio. (b) Profiles of temperature, heat release rate, and equivalence ratio across the laminar flame with $\phi$ = 0.4. The spatial coordinate is normalized by the flame thickness for DC PD, $\delta_{L\,\mathrm{ref}}^0$, and the origin is located at the position of maximum heat release rate.}
\label{Fig_1DFlame}
\end{figure}

An initial comparison of the three versions of the TC model is performed in a one-dimensional (1D) freely propagating premixed flame to help explain their behavior in the turbulent flame. The evolution of the laminar flame speed and thermal flame thickness with the equivalence ratio is shown in Figure~\ref{Fig_1DFlame}(a). Results obtained with detailed chemistry using the chemical kinetics simulation tool Cantera \cite{GoodwSpeth21} are included as a reference, considering unity Lewis number (DC Le1) and including preferential and differential diffusion (DC PD).  

The results show a thinner flame when a unity Lewis number is assumed than when preferential and differential diffusion are considered. This is known to result from the increased diffusion in the latter case, which leads to a broader distribution of species, radicals, and thus the heat release rate over a wider region of the flame front \cite{AspdeDay11b, BergeAttil22}. On the other hand, the effect of the unity Lewis number assumption on the laminar flame speed varies with equivalence ratio as a result of the interplay between chemical kinetics, species and thermal diffusion, and thermodynamics \cite{Pitsc24}. For equivalence ratios above 0.6, the unity Lewis number assumption underpredicts the laminar flame speed because it neglects the enhanced diffusion of reactants into the flame and of radicals from the flame toward the unburned mixture ahead. This diffusion enhances the reaction rate and increases the flame speed, even though it makes the mixture locally leaner. For equivalence ratios below 0.6, the relatively low flame temperature makes the reaction rate highly sensitive to the radical concentration. The reduced diffusion under the unity Lewis number assumption leads to higher radical concentrations in the reaction zone and therefore to an overprediction of the reaction rate and the laminar flame speed for very lean mixtures.

The TC PD model accurately predicts the laminar flame speed and thermal flame thickness in agreement with detailed chemistry \cite{PerezForte25}. In contrast, when preferential and differential diffusion are considered only in the flamelet tabulation (TC PD-F), the laminar flame speed is consistently overpredicted across the entire equivalence ratio range. For an equivalence ratio of 0.4, the laminar flame speeds predicted by the TC Le1, TC PD-F, and TC PD models are 0.393, 0.660, and 0.206 m/s, respectively.


The profiles of temperature, heat release rate, and equivalence ratio across the laminar flame with $\phi = 0.4$ are shown in Figure~\ref{Fig_1DFlame}(b). The detailed chemistry with preferential and differential diffusion exhibits a slower temperature rise with a broader preheating zone compared to the unity Lewis number assumption, due to the broader and less intense heat release distribution and the local decrease in equivalence ratio due to the differential diffusion \cite{ChenIm00, AspdeDay11b}. While the TC PD model accurately captures this flame structure, the TC PD-F version does not represent the actual species transport, and the local equivalence ratio remains constant, artificially maintaining higher reactivity and leading to a more intense heat release rate. In other words, the TC PD-F model accesses the state in the manifold corresponding to richer mixtures.

\begin{figure}[h!]
\centering
\includegraphics[trim = 0 105 730 0 ,clip, width=0.5\linewidth]{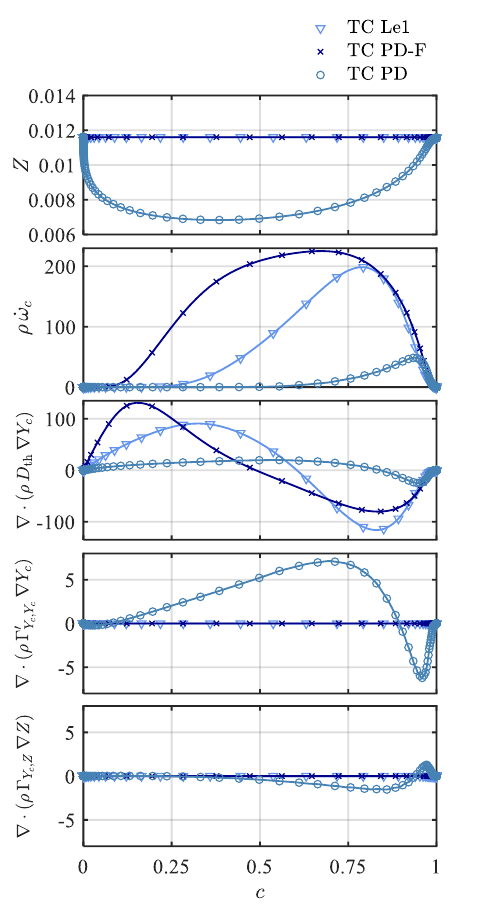}
\caption{Mixture fraction and transport budget of progress variable for the unstretched hydrogen-air laminar flame with $\phi$ = 0.4.}
\label{Fig_1DFlame_budget}
\end{figure}

The behavior of the three versions of the TC model can also be analyzed from the budget of the different terms in the progress variable transport equation, as shown in Figure~\ref{Fig_1DFlame_budget} along the laminar flame in the normalized progress variable coordinate. The evolution of the mixture fraction is also included. The source term provides the largest contribution among the different terms. On the one hand, comparing TC Le1 and TC PD-F highlights the effect of preferential and differential diffusion at the thermochemical level, which increases the source term due to enhanced diffusion of fuel (H$_2$) into the flame and of radicals ahead of the flame in the flamelet. On the other hand, comparing TC PD-F and TC PD highlights the effect of preferential and differential diffusion at the transport level, which reduces the source term due to the local decrease in mixture fraction caused by preferential diffusion.

The diffusion terms also contribute to flame propagation by controlling how rapidly the progress variable spreads ahead of the flame. The normal diffusion term exceeds the source term in the preheating zone, where flame propagation is governed by diffusion. Overall, the lowest value of the diffusion terms corresponds to the TC PD model, as the reduced source term results in a less steep progress variable gradient. To isolate the contribution of differential diffusion from the unity Lewis number component in the normal diffusion term, $\Gamma_{Y_c,Y_c}$ is decomposed into $D_{\mathrm{th}}$ and $\Gamma_{Y_c,Y_c}' = \Gamma_{Y_c,Y_c} - D_{\mathrm{th}}$. The contribution of differential diffusion to the normal diffusion term is approximately one-fourth of the unity Lewis number contribution.

The normal diffusion terms induce diffusion of the progress variable in the direction opposite to its gradient, resulting in diffusion in the same direction as flame propagation (from right to left in Figure~\ref{Fig_1DFlame_budget}). In contrast, the cross-diffusion term, associated with $\Gamma_{Y_c,Z}$, induces diffusion of the progress variable mainly in the direction opposite to flame propagation. This opposing diffusion contributes to a reduction in flame speed for the TC PD model compared to TC PD-F. However, its contribution remains small compared to that of the normal diffusion.

The budget for the mixture fraction control variable is not included for brevity. The local decrease in mixture fraction (equivalence ratio) caused by preferential diffusion is governed by the cross-diffusion term in the mixture fraction transport equation, which induces diffusion of $Z$ from the preheating zone toward the reaction zone. This leaner mixture fraction region is convected downstream while being smoothed by the normal diffusion term. The convective–diffusive balance results in the mixture fraction recovering the value of the global mixture fraction toward the end of the laminar flame, as shown at the top of Figure~\ref{Fig_1DFlame_budget}.


\subsection{Turbulent premixed flame}

\begin{figure}[h!]
\centering
\includegraphics[trim = 0 75 630 10 , clip, width = 0.55\textwidth]{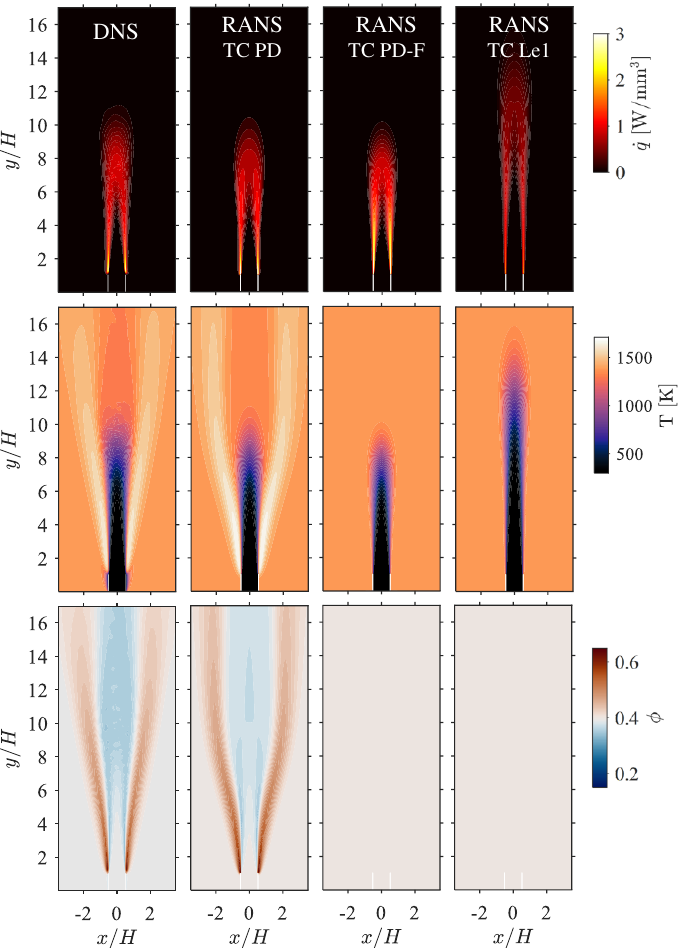}
\caption{Favre-averaged DNS and RANS contours of heat release rate, temperature, and equivalence ratio for the hydrogen-air slot flame with Re = 5500.}
\label{Fig_DNSvRANS_Re5500}
\end{figure}

\begin{figure}[h!]
\centering
\includegraphics[trim = 0 105 630 10 , clip, width = 0.55\textwidth]{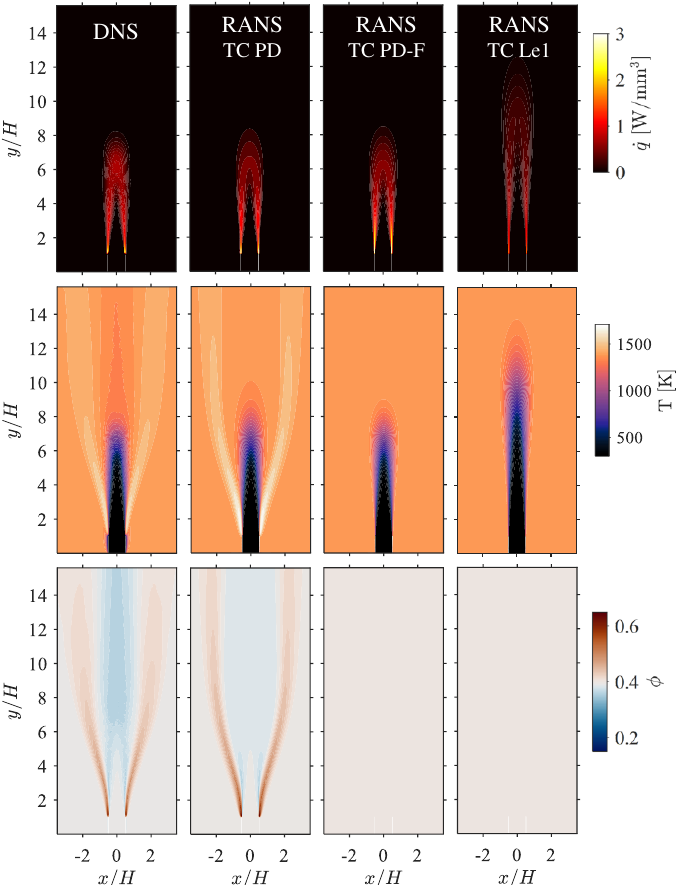}
\caption{Favre-averaged DNS and RANS contours of heat release rate, temperature, and equivalence ratio for the hydrogen-air slot flame with Re = 11000.}
\label{Fig_DNSvRANS_Re11000}
\end{figure}

The results of the simulations of the turbulent hydrogen\--air slot flame are presented in Figures~\ref{Fig_DNSvRANS_Re5500} and~\ref{Fig_DNSvRANS_Re11000} for Reynolds numbers of 5500 and 11000, respectively. The contours of heat release rate, temperature, and equivalence ratio obtained from the RANS simulations are compared with DNS data, first Favre-averaged and then spatially averaged along the spanwise direction.

The DNS Favre-averaged fields show local variations of the equivalence ratio that lead to regions of super-adiabatic temperature where the mixture is richer and under-adiabatic temperature where it is leaner. As mentioned before, the fluctuations in the equivalence ratio along the instantaneous flame front (shown in Figure~\ref{Fig_DNS_flame}) are linked to flame stretch in the form of curvature and strain rate \cite{BergeAttil22,BergeAttil24}. It is known that the misalignment between diffusion and convection at the flame front, induced by either curvature or strain rate, leads to local variations in the equivalence ratio for mixtures with non-unity Lewis number \cite{ChungLaw88, Law89}. In addition to the curvature and strain rate induced by turbulence and thermodiffusive instabilities, the slot flame is exposed to high levels of strain rate in the shear layer at the burner outlet \cite{BergeAttil24}. In this region, the strong misalignment between diffusion, normal to the flame front, and convection along the streamwise direction leads to an increase in the equivalence ratio behind the flame front associated with the mean positive strain rates \cite{BergeAttil22}, similar to that observed in laminar counterflow premixed flames \cite{BergeAttil24, PorcaLange24}.

The RANS simulations with the TC PD model capture both the flame length and the heat release distribution along the flame brush. Moreover, the temperature field shows that the RANS simulations with the TC PD model are able to predict the super-adiabatic regions at the shear layer of the slot flame. The peak temperature at the flame base is slightly overpredicted, as heat losses to the burner walls are neglected. In general, the agreement is better near the flame base, close to the burner outlet, whereas toward the flame tip the RANS simulations predict a smaller decrease in the equivalence ratio and, consequently, in the temperature than the DNS. Toward the flame tip, turbulence levels are higher, and unresolved turbulence–diffusion interactions may play a more significant role. Nevertheless, the key characteristics of the turbulent hydrogen-lean flame are, in general, well captured by the RANS simulations with the TC PD model and good quantitative agreement is obtained, as shown in Section~\ref{S:TurbDiff}. Furthermore, the inclusion of enthalpy as a control variable is expected to improve the accuracy of the TC PD model by better describing the cross-diffusion due to preferential and differential diffusion \cite{FortePerez25,SchepvanO25}.


Since preferential and differential diffusion are included only at the thermochemical level and not at the transport level, the RANS simulation with the TC PD-F model does not predict regions of equivalence ratio variation or super-adiabatic temperature. On the one hand, although the flame length predicted by the TC PD-F model is slightly shorter than that obtained with the TC PD model, the difference remains modest compared with the substantially larger disparity in laminar flame speed, which is more than three times greater for the TC PD-F model. On the other hand, the RANS simulations predict a much longer flame when a unity Lewis number is assumed (TC Le1 model), despite the laminar flame speed being higher than for the TC PD model. This agrees with DNS results using a unity Lewis number \cite{BergeAttil22} and is explained by the higher overall burning rate of the turbulent flame when preferential and differential diffusion are considered, associated with increased flame surface area and higher average local burning rates due to predominantly positive stretch \cite{BergeAttil22,BergeAttil24}. Since flame front dynamics are not resolved in the RANS simulations, the differences in the predicted flame length among the TC models must be explained from the resolved mean fields and turbulence\-–chemistry interaction.

\begin{figure}[h!]
\centering
\includegraphics[trim = 0 75 300 0 , clip, width = 0.95\textwidth]{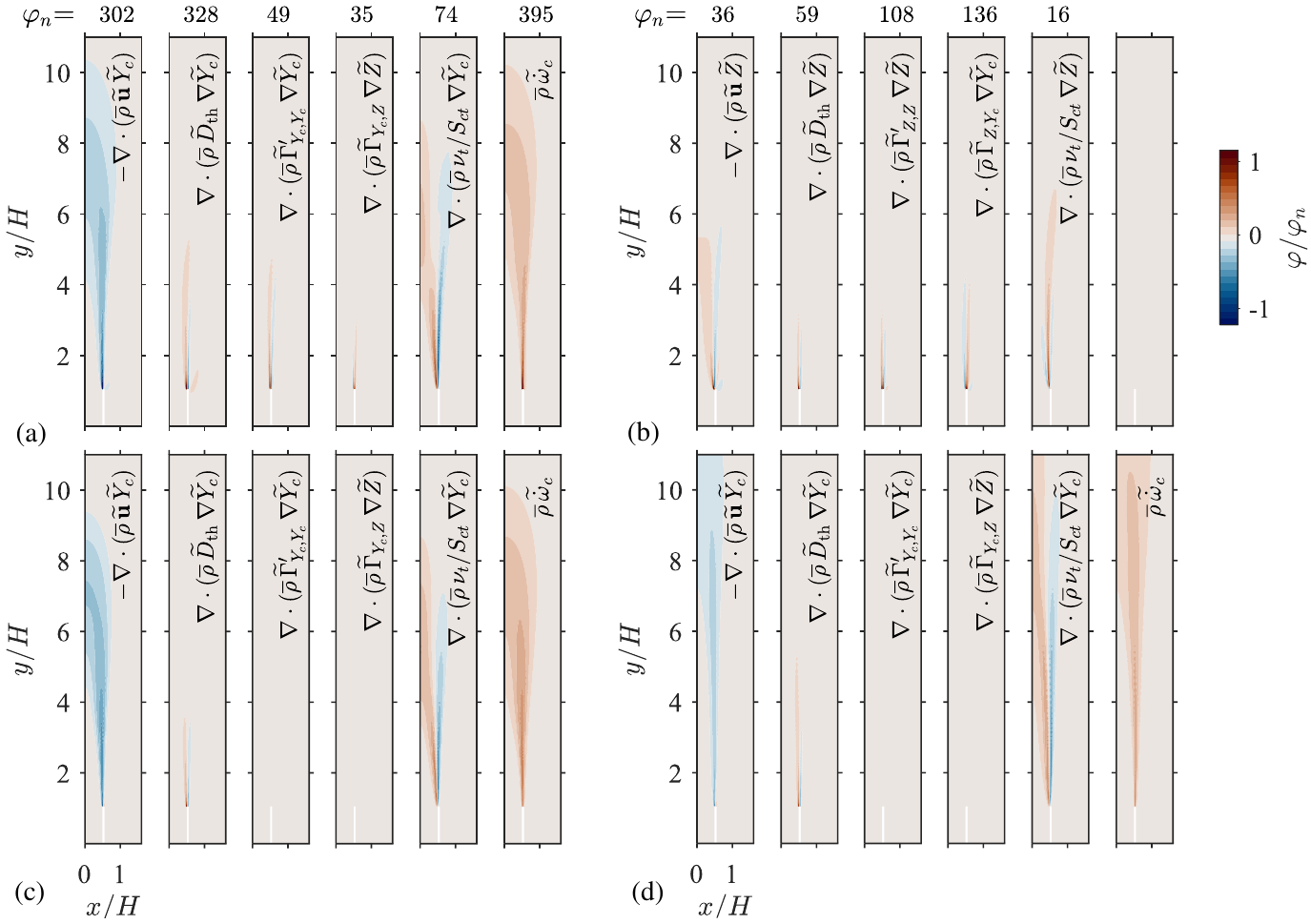}
\caption{Contours of the RANS scalar transport budget for the hydrogen-air slot flame at Re = 5500: (a) progress variable and (b) mixture fraction for TC PD, (c) progress variable for TC PD-F, and (d) progress variable for TC Le1, where $\varphi$ represents the various terms in the control variable transport equations and $\varphi_n = P_{99}(|\varphi|_{>0})$. Fields in (c) and (d) are normalized with the same $\varphi_n$ values as in (a).}
\label{Fig_budget_contour}
\end{figure}

As for the laminar flame, a budget analysis of the control variable transport equations is performed to understand the results obtained with the three versions of the TC model in the RANS simulation. The contours of each transport term in the progress variable and mixture fraction equations are presented in Figure~\ref{Fig_budget_contour} for the case with a Reynolds number of 5500 to provide a visual picture of their spatial distribution. To improve visualization, each transport term is normalized by the 99$^{\text{th}}$ percentile of non-zero values for the corresponding term in the TC PD case. All diffusion terms have their highest contributions (peak values) at the flame base, closer to the burner outlet, where gradients are steeper, and decrease downstream, where the gradients are spread out by turbulence. Similarly, the mean reaction zone, represented by the progress variable source term, is concentrated in a narrow region resembling a flame sheet at the shear layer, while it spreads over a wide flame brush at the flame tip.

From the transport terms in the mixture fraction transport equation for the TC PD model in Figure \ref{Fig_budget_contour}(b), it can be seen that the local variation of the mixture fraction primarily occurs in the shear layer at the flame base and then these variations are convected downstream while being dissipated by the molecular and turbulent diffusion terms, leading to the equivalence ratio distribution shown in Figures~\ref{Fig_DNSvRANS_Re5500} and \ref{Fig_DNSvRANS_Re11000}. Similar behavior occurs for the case with a Reynolds number of 11000 (not shown for brevity); however, as turbulence increases, the strength of the cross-diffusion term decreases, leading to a reduction in local variation of the mixture fraction and, consequently, in the values of the super-adiabatic temperature shown in Figure~\ref{Fig_DNSvRANS_Re11000}.

\begin{figure}[h!]
\centering
\includegraphics[trim = 0 140 330 0  , clip, width = 1.1\textwidth]{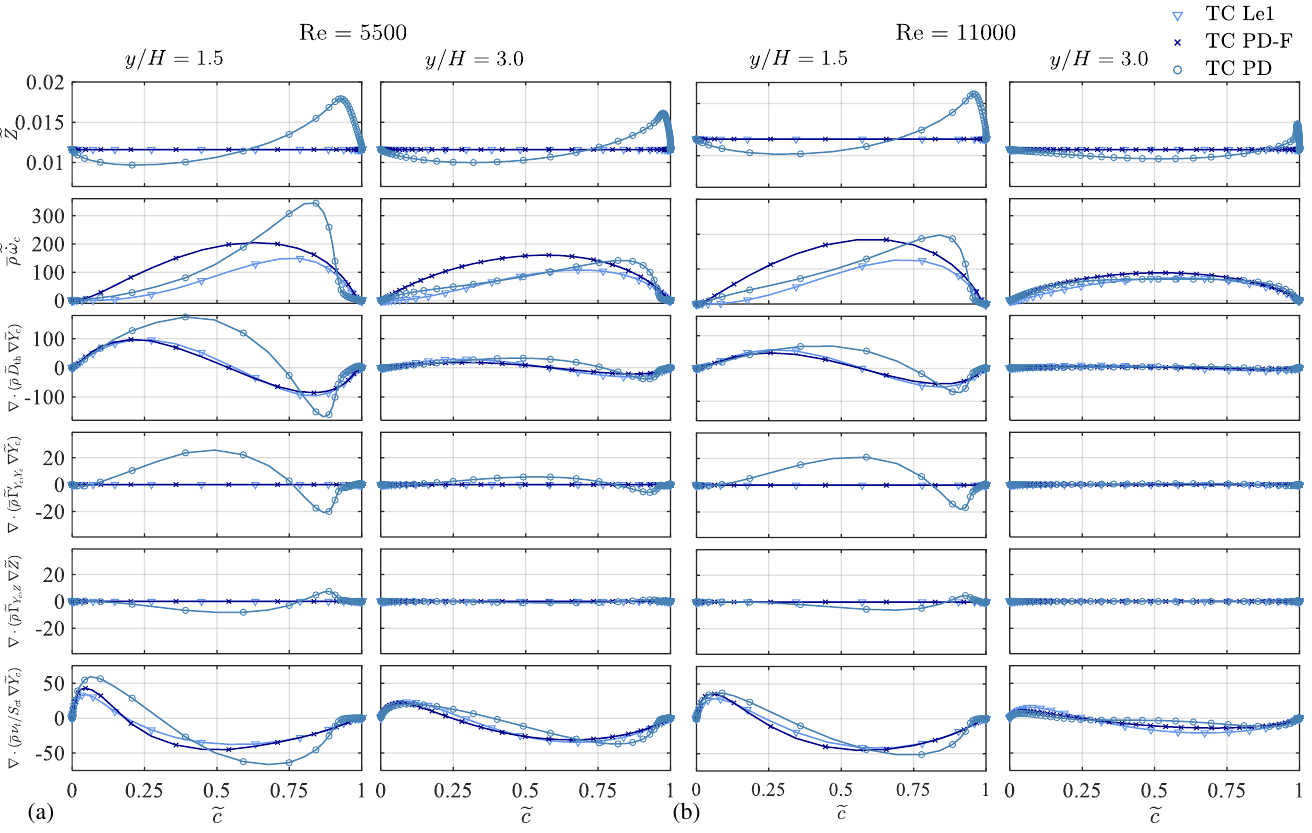}
\caption{Mixture fraction and transport budget of the progress variable for the slot flame with (a) Re = 5500 and (b) Re = 11000 at $y/H$ = 1.5 and 3.0.}
\label{Fig_budget_c_profile}
\end{figure}

A quantitative comparison of the budget of the progress variable transport equation and the mixture fraction across the mean flame sheet as a function of the normalized progress variable is presented in Figure~\ref{Fig_budget_c_profile} at the heights $y/H = 1.5$ and 3.0. As described earlier, preferential diffusion leads to variations in the mixture fraction due to the mean positive strain rate in the shear layer, with a decrease ahead of the flame front and an increase toward the end of the reaction zone. The evolution of the mixture fraction in the turbulent flame with the TC PD model leads to a different evolution of the progress variable source term in the RANS simulations compared to the laminar flame. The source term for the TC PD model now remains consistently higher than that of the TC Le1 model and higher than for the TC PD-F model where the mixture is richer and lower where it is leaner.

The increase in the progress variable source term in the RANS simulation with the TC PD model, leads to steeper progress variable gradients and thus to a higher contribution of the diffusion terms compared to the 1D laminar flame. However, the relative strength of the various components of the molecular diffusion remains similar to the laminar case. The contribution of differential diffusion to the normal diffusion term still represents around 1/5 of the total normal diffusion, while the cross-diffusion is approximately 1/20 of the normal diffusion.  The contribution of turbulence to the spreading of the control variables in the RANS simulation is represented by the turbulent diffusion term, which is proportional to $\rho\nu_t/S_{ct}$. The turbulent diffusion term is lower than the molecular diffusion terms at the flame base near the burner, while it becomes dominant towards the flame tip.

\begin{figure}[h!]
\centering
\includegraphics[trim = 0 345 750 0 , clip, width = 0.5\textwidth]{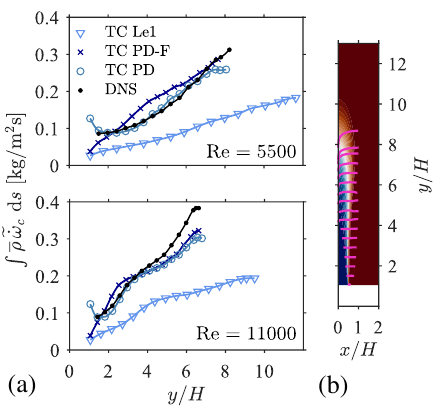}
\caption{(a) Integral of the progress variable source term $\overline{\rho} \, \widetilde{\dot{\omega}}_c$ along flame trajectories $s$. The streamwise position $y/H$ of each trajectory is referenced to the location of $\tilde{c}$ = 0.5 in the trajectory. (b) Example of the flame trajectories $s$ (pink lines) overlaid on the normalized progress‐variable field $\tilde{c}$ for Re = 5500 with the TC PD model.}
\label{Fig_int_c_s}
\end{figure}

Both diffusion and source terms contribute to the rate at which the combustion process progresses in the RANS simulation; however, it is the source term that primarily governs the propagation of the turbulent flame and determines the flame front location. The evolution of the progress variable source term from the flame base to the flame tip is shown in Figure~\ref{Fig_int_c_s} by its integral along trajectories $s$ across the mean flame front. These trajectories are orthogonal to the $\widetilde{c}$ isolines, following the streamlines of the normalized progress variable gradient ($\nabla \widetilde{c} / \vert{}\nabla \widetilde{c}\vert{}$) from the unburned reactants to the fully burned products. This integral can be interpreted as the reaction rate normal to the mean flame front, which is related to the rate at which the fuel is consumed. The DNS data for $\overline{\rho} , \widetilde{\dot{\omega}}_{H2O}$ are included as a reference.

The reaction rate is lower across the entire shear layer when using the TC Le1 model, resulting in a longer flame and a wider flame brush. Compared to the TC PD-F model, the TC PD model exhibits a higher reaction rate at the flame base, where rich mixture zones lie within the reaction zone. As these rich mixture zones are convected downstream and out of the reaction zone, the slightly leaner mixture left behind results in a lower reaction rate. In the RANS simulations, the discrepancy between the TC PD and TC PD-F models decreases downstream and with increasing Reynolds number as the mixture fraction variation decreases. This explains why accounting for preferential and differential diffusion only at the thermochemical level yields a similar prediction of the flame length.

\subsection{Turbulence\--chemistry interaction}
\label{S:TCI}

\begin{figure}[ht]
\centering
\includegraphics[trim = 0 80 632 0 , clip, width = 0.55\textwidth]{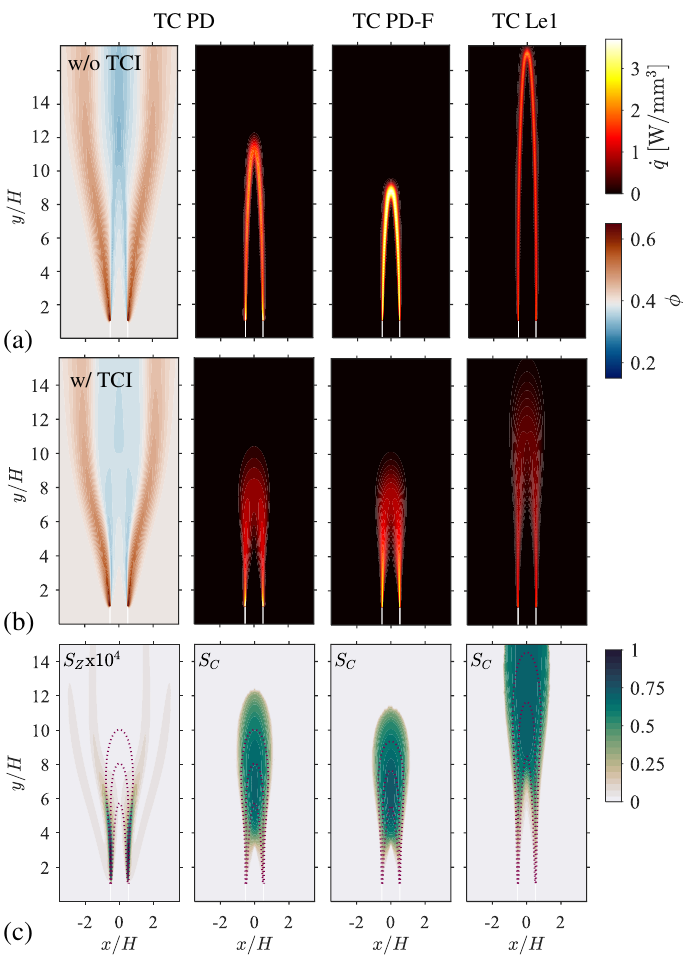}
\caption{Contours of equivalence ratio and heat release rate from the RANS simulations of the hydrogen-air slot flame ($Re = 5500$): (a) without turbulent-chemistry interaction, where $S_C = S_Z = 0$, (b) with turbulent-chemistry interaction, and (c) the corresponding normalized variances $S_C$ and $S_Z$. Dotted lines correspond to isolines of $\tilde{c}$ equal to 0.1, 0.5, and 0.9.}
\label{Fig_RANS_noVar_Re5500}
\end{figure}

Toward the flame tip, turbulence spreads the reaction zone into a wide flame brush, where most of the fuel is consumed. As a result, the flame structure and fuel consumption rate are primarily governed by the turbulence\--chemistry interaction. The influence of TCI is illustrated in Figure~\ref{Fig_RANS_noVar_Re5500}, which presents additional RANS results obtained by suppressing TCI for the three versions of the TC model. To suppress TCI, the variances of the mixture fraction and progress variable are set to zero before accessing the turbulent manifold, effectively reducing the model to the laminar manifold. As a result, the RANS simulation yields a highly intense reaction zone concentrated in a sheet-like flame front, rather than forming a flame brush. The flames without TCI are longer than those with TCI, especially when a unity Lewis number is assumed. The difference between the TC PD and TC PD-F models is more pronounced when TCI is suppressed. The more concentrated reaction zone of the flame without TCI results in stronger diffusion due to the higher gradients across the flame, which leads to greater variation of the equivalence ratio.

Figure \ref{Fig_RANS_noVar_Re5500}(c) shows the distributions of the normalized progress variable variance and mixture fraction variance in the RANS simulations with TCI. While the mixture fraction variance for the TC PD case is technically non-zero, its magnitude ($\mathcal{O}(10^{-5})$) is negligible. This indicates that only three control variables, $Y_c$, $Z$, and $Y_{c,v}$, are needed to describe the fully premixed turbulent slot flame, in agreement with previous analyses \cite{BergeAttil25}. In contrast, the progress variable variance exhibits high values, peaking at the flame tip. For the TC PD case at Re = 5500, the peak and volume-averaged $S_C$ values are 0.73 and 0.34, respectively. These values increase to 0.84 and 0.64 for Re = 11000. While the TC PD-F model yields a similar distribution, the TC Le1 case produces slightly higher values, as its longer flame is exposed to stronger turbulence downstream. 

\begin{figure}[h!]
\centering
\includegraphics[trim = 0 240 625 0 , clip, width = 0.55\textwidth]{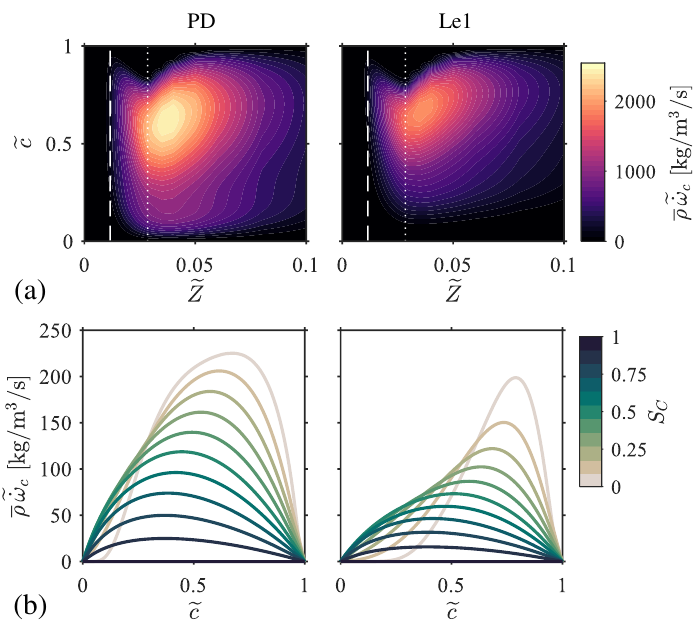}
\caption{Progress variable source term for the tables with PD (left column) and Le1 (right column). (a) Contours over mixture fraction $Z$ and normalized progress variable $c$ for $S_C = S_Z$ = 0. The dotted line indicates $\phi = 1.0$ and the dashed line indicates $Z = 0.0116$ $(\phi = 0.4)$.
(b) Profiles along $c$ for $Z = 0.0116$ and $S_Z$ = 0 with varying normalized progress variable variance $S_C$.}
\label{Fig_Table_sourceYc_Sc}
\end{figure}

The effect of preferential and differential diffusion at the thermochemical level can be seen directly in the manifold. Figure~\ref{Fig_Table_sourceYc_Sc}(a) shows the evolution of the progress variable source term in the laminar manifold as a function of the normalized progress variable and mixture fraction. When preferential and differential diffusion are considered instead of the unity Lewis number assumption, the source term increases throughout the $Z$–$c$ space, especially at rich mixture fractions. Preferential and differential diffusion also spread the progress variable source term over a wider range of progress variable values toward the unburned condition. This is related to the differential diffusion of highly diffusive radicals ahead of the flame front in the laminar flamelets, causing an earlier ignition of reaction at correspondingly lower values of $c$ \cite{Pitsc24}. 

As previously described, TCI is accounted for by integrating the laminar manifold with $\beta$-shaped PDFs, resulting in a turbulent manifold. Figure~\ref{Fig_Table_sourceYc_Sc}(b) shows the evolution of the progress variable source term in the turbulent manifold as a function of the normalized progress variable and normalized progress variable variance along the constant mixture fraction corresponding to $\phi = 0.4$. Note that, while for the TC Le1 and TC PD-F models this corresponds to the evolution of the source term along the flame front, this is not the case for the TC PD model due to the variation of the mixture fraction, as seen in Figures~\ref{Fig_1DFlame_budget} and~\ref{Fig_budget_c_profile}.

The effect of turbulence on the reaction rate can be seen in the evolution of the progress variable source term at various levels of normalized progress variable variance, $S_C = c_v/\widetilde{c} \left(1 - \widetilde{c} \right)$, ranging from 0, indicating no fluctuations (laminar flamelet), to 1, indicating the maximum possible fluctuation, where the mixture alternates between fresh and fully burned gases, driven solely by turbulent mixing, with no partially reacted states. Increasing the progress variable variance spreads the source term over $\widetilde{c}$ while decreasing its peak value. The interplay between the reduction in the local reaction rate and the broadening of the reaction zone governs the TCI in the tabulated chemistry model. The spreading of the source term over $\widetilde{c}$ allows the reaction zone to extend in physical space across a wider flame brush thickness, representing the increased flame surface area due to turbulence-induced wrinkling. Consequently, the overall reaction rate increases because a larger region participates in the reaction, even though the local reaction rate decreases. As the variance increases further under strong turbulence, the source term continues to decrease until it eventually vanishes. Compared to the unity Lewis number assumption, preferential and differential diffusion already spreads the progress variable source term over a wide range of $\widetilde{c}$ in the absence of turbulence. Hence, the effect of turbulence, i.e. the progress variable variance, further enhances this spreading. 



Finally, it is worth mentioning that one effect of preferential and differential diffusion in lean premixed hydrogen flames is the formation of cellular structures at the flame front due to thermo-diffusive instabilities \cite{Pitsc24}. These instabilities may interact synergistically with turbulence, further increasing the global reaction rate by enhancing both the flame surface area and the local reaction rates \cite{BergeAttil22}. Currently, this effect is not explicitly included in the TCI and should be investigated further.

\subsection{Turbulent diffusion}
\label{S:TurbDiff}

The TC model with preferential and differential diffusion has already shown the ability to capture the key characteristics of the lean hydrogen combustion in the RANS simulation of the turbulent slot flame. However, its quantitative performance is affected by how the different aspects of turbulent transport are modeled in the RANS simulation.

Turbulent diffusion models the transport of scalars by turbulent fluctuations, not resolved in the Favre-averaged transport equations. The gradient diffusion hypothesis is used to model the turbulent scalar fluxes, with an effective diffusivity defined through the turbulent Schmidt number. A lower $Sc_{t}$ increases the effective diffusion, thereby enhancing the turbulent transport of $Y_c$ and $Z$. This results in stronger mixing of the mixture fraction and increased spreading of the progress-variable field. No single universal constant value of $Sc_{t}$ has been established, and in practice, turbulence models rely on empirically chosen values for different situations \cite{TominStath07}. While values in the range of 0.4–0.9 are commonly used in combustion simulations \cite{JiangCampb09,ShiChen16}, more accurate predictions have been reported for hydrogen combustion with lower values of $Sc_{t}$, in the range of 0.2–0.5 \cite{PoschPfitz08,AkbarHill11,LopezSun20}.

\begin{figure*}[h!]
\centering
\includegraphics[trim = 0 5 325 0  ,clip,width=1.1\textwidth]{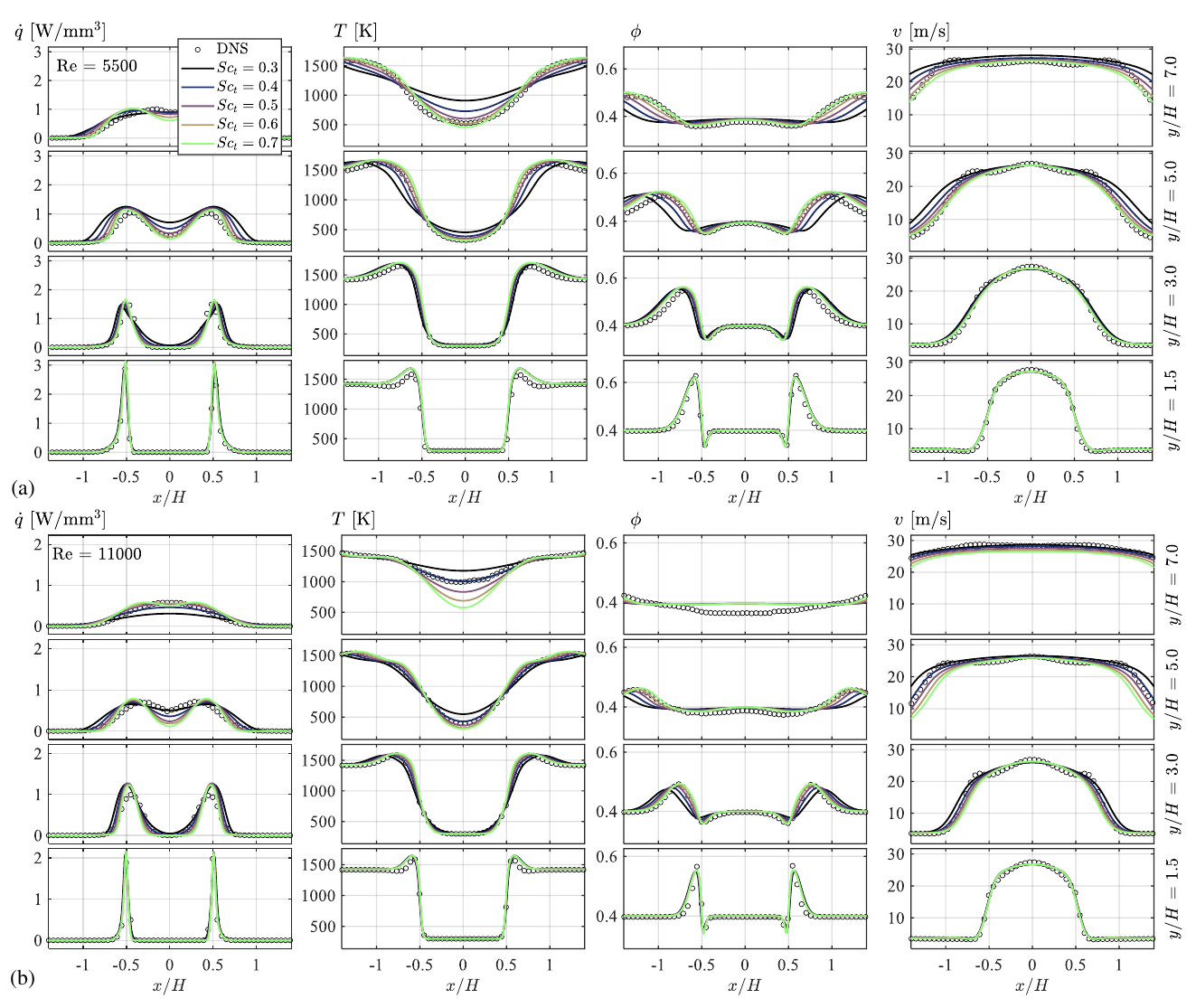}
\caption{Profiles of heat release rate, temperature, equivalence ratio, and $y$-velocity for the hydrogen-air slot flame with (a) Re = 5500 and (b) Re = 11000 at $y/H$ = 1.5, 3.0, 5.0, and 7.0. Favre-averaged DNS and RANS with varying $Sc_{t,c}$.}
\label{Fig_DNSvRANS_profiles_Sct}
\end{figure*}

A common assumption is that turbulent eddies mix all scalars at the same rate, regardless of their molecular diffusivities; therefore, a single turbulent Schmidt number can be applied to all transported scalars. However, this assumption is not necessarily valid for the mixture fraction and the progress variable, since they represent different physical processes. In the present case, $Sc_{t,z}$ does not have a significant impact on the results because variations in the mixture fraction are relatively small; therefore, its effect cannot be assessed independently, and it is set equal to $Sc_{t,c}$. A more comprehensive evaluation of the impact of independent turbulent Schmidt numbers should be conducted for non-premixed or partially premixed cases, where mixing and reaction progress interact more strongly.

Figure~\ref{Fig_DNSvRANS_profiles_Sct} shows the results of the RANS simulations using the TC PD model with $Sc_{t,c}$ varying from 0.3 to 0.7. The value of $Sc_{t,c}$ affects the various fields, especially toward the flame tip, where the role of turbulent diffusion is more pronounced, while its effect is minimal at the flame base, where molecular diffusion dominates, as shown in Figure~\ref{Fig_budget_c_profile}. Moving downstream along the flame height, the gradients are smoothed out as turbulent diffusion increases with lower $Sc_{t,c}$, leading to reduced equivalence ratio variation. 

\begin{figure}[h!]
\centering
\includegraphics[trim = 0 330 625 5 , clip, width = 0.55\textwidth]{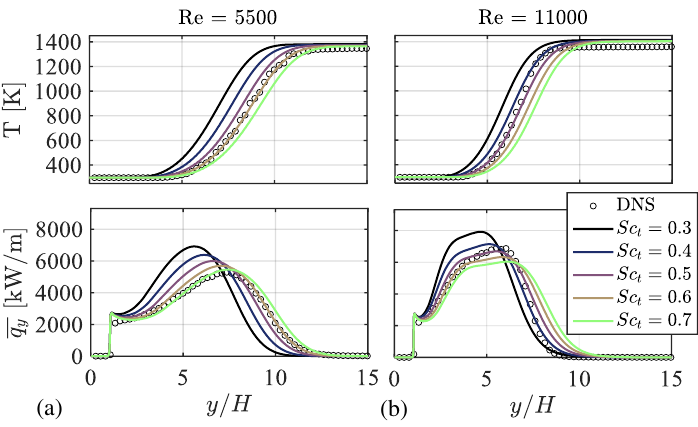}
\caption{Profiles of temperature (at $x/H$ = 0.0) and streamwise distribution of the integrated heat release rate for the hydrogen-air slot flame with (a) Re = 5500 and (b) Re = 11000. Favre-averaged DNS and RANS with varying $Sc_t$}
\label{Fig_DNSvRANS_axis_Sct}
\end{figure}

The impact on the flame length is evaluated in Figure~\ref{Fig_DNSvRANS_axis_Sct} through the axial profiles of temperature and the streamwise heat release rate distribution, given by $\overline{q_y} = \int{\dot{q} (x,y) }\dd x$. The temperature rise is delayed and its slope is slightly reduced with higher turbulent Schmidt numbers, indicating a decrease in the overall turbulent consumption speed and a downstream spread of the reaction zone. This becomes clearer when examining the heat release rate distribution $\overline{q_y}$, where the reaction zone shifts downstream, spreading out and lowering its peak value as turbulent diffusion decreases. The turbulent Schmidt number affects not only the diffusion of the progress variable and mixture fraction but also the diffusion and production of their variances, thereby influencing turbulence\--chemistry interaction as well. 

It can be seen from Figures~\ref{Fig_DNSvRANS_profiles_Sct} and \ref{Fig_DNSvRANS_axis_Sct} that the value of $Sc_{t,c}$ that best agrees with the DNS data for the slot flame is not the same for the two Reynolds numbers evaluated, where higher values, around 0.6, are in better agreement for the Re = 5500 case and lower values, around 0.4, for the Re = 11000 case. Although a constant value of 0.5 was selected for the RANS simulations of the slot flame burner in the present study, these results indicate that further work is needed on the modeling of turbulent diffusion in RANS simulations of premixed hydrogen flames.

\subsection{Turbulent scalar dissipation rate}
\label{S:TurbSDR}

The turbulent scalar dissipation rate is the rate at which the fluctuating components of scalar fields, such as the mixture fraction and the progress variable, are irreversibly smoothed out by molecular diffusion once they have been strained down to the smallest turbulent scales. In analogy to how turbulent kinetic energy is dissipated by viscosity into heat, the scalar dissipation rate quantifies how fast scalar variances are dissipated into the mean field \cite{PoinsVeyna05}. 

\begin{figure*}[h!]
\centering
\includegraphics[trim = 0 5 325 0  ,clip,width=1.1\textwidth]{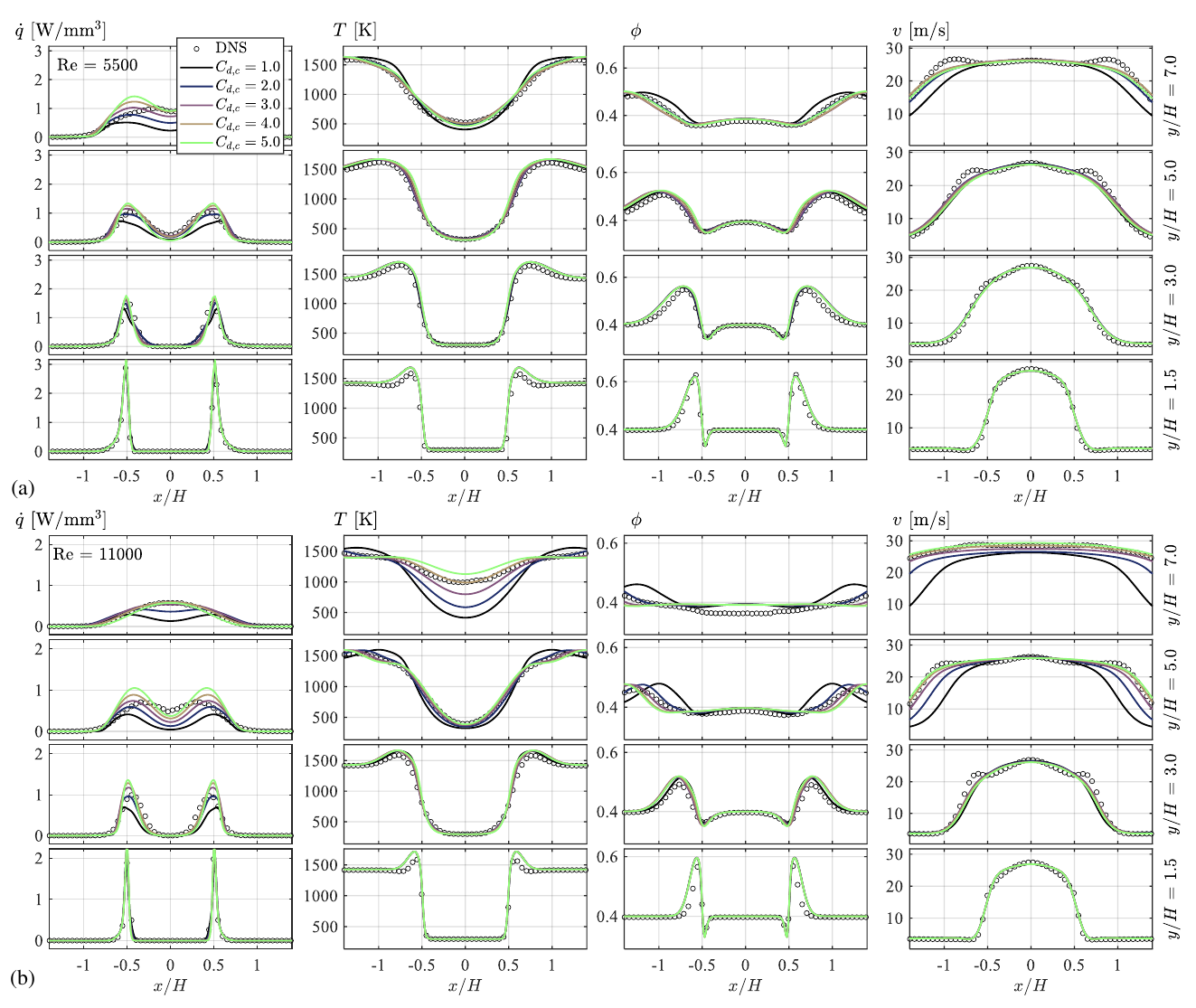}
\caption{Profiles of heat release rate, temperature, equivalence ratio, and $y$-velocity for the hydrogen-air slot flame with (a) Re = 5500 and (b) Re = 11000 at $y/H$ = 1.5, 3.0, 5.0, and 7.0. Favre-averaged DNS and RANS with varying $C_{d,c}$.}
\label{Fig_DNSvRANS_profiles_Cd}
\end{figure*}

\begin{figure}[h!]
\centering
\includegraphics[trim = 0 330 625 5 , clip, width = 0.55\textwidth]{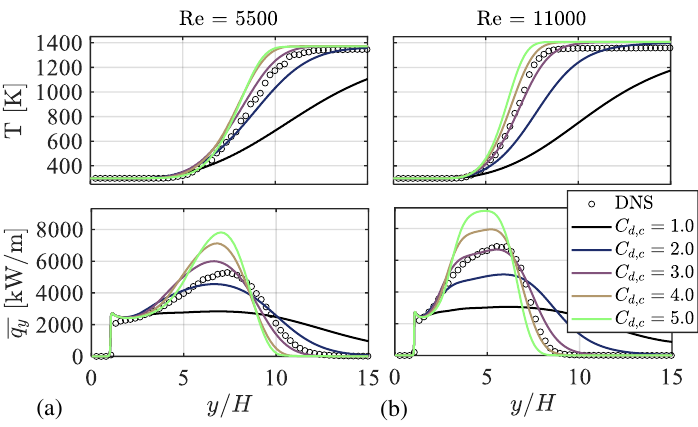}
\caption{Profiles of temperature ($x/H$ = 0.0) and streamwise distribution of the integrated heat release rate for the hydrogen-air slot flame with (a) Re = 5500 and (b) Re = 11000. Favre-averaged DNS and RANS with varying $C_{d,c}$.}
\label{Fig_DNSvRANS_axis_Cd}
\end{figure}

Since turbulence is not resolved in the RANS simulation, the rate at which variance is dissipated is modeled as proportional to the ratio between the scalar variance and the turnover time of the smallest eddies, with the proportionality given by the scalar dissipation constants, $C_{d}$. Some studies recommend values in the range of 0.25–2 depending on the mixing regime, and similar to the turbulent Schmidt number, different values may be used for the mixture fraction variance and the progress variable variance \cite{PoinsVeyna05}.

Figure~\ref{Fig_DNSvRANS_profiles_Cd} shows the results of the simulations using the TC PD model with $C_{d,c}$ varying from 1.0 to 5.0. Since the effect of the mixture fraction variance is negligible in the fully premixed flame, $C_{d,z}$ has no impact and is therefore set equal to $C_{d,c}$ in the current evaluation. The turbulent dissipation rate increases with $C_{d,c}$; hence, the reaction zone is subjected to lower levels of progress variable variance. As $C_{d,c}$ increases, the flame brush contracts into a thinner, more sheet-like front with a sharper heat release rate. The thinning of the flame brush is also reflected in the velocity field as locally enhanced acceleration (due to velocity–density coupling), which pushes the rich, super-adiabatic branches at the shear layer away from the flame axis. The impact of $C_{d,c}$ is greater where turbulence is stronger, either downstream in the flame, where progress-variable variances are high, or at higher Reynolds numbers.

The effect of the scalar dissipation constant on the flame length is shown in Figure~\ref{Fig_DNSvRANS_axis_Cd}. Its influence is small near the flame base, whereas toward the tip the downstream spreading of the reaction zone, and thus the effective flame length, changes significantly with $C_{d,c}$. This behavior results from the increase in the progress variable variance as $C_{d,c}$ decreases and its impact on the flame brush through turbulence--chemistry interaction, as discussed in Section~\ref{S:TCI}.

\subsection{Turbulence model}
\label{S:TurbModel}

The accuracy of RANS simulations strongly depends on the turbulence model used to close the Favre-averaged Navier–Stokes equations. The turbulence model dictates how the unresolved Reynolds stresses are represented, and thus directly affects turbulence\--chemistry interaction, turbulent diffusion, and the scalar dissipation rate. Consequently, the predictive capabilities of the tabulated chemistry model with preferential and differential diffusion are inherently limited by the accuracy of the RANS turbulence model. In this section, the impact of different turbulence models on the RANS simulation of the turbulent slot flame is evaluated. 

Three additional turbulence models are considered: the realizable $\kappa$–$\epsilon$ model \cite{ShihLiou95}, the $\kappa$–$\omega$ SST (Shear Stress Transport) model \cite{Mente94}, which blends the standard $\kappa$–$\omega$ formulation near the wall with the $\kappa$–$\epsilon$ model in the far field, and the baseline Reynolds stress model (RSM-BSL) \cite{Wilco98}, which solves transport equations for all Reynolds-stress components and thereby captures turbulence anisotropy and nonlinear effects. Default model constants are retained \cite{ANSYS23} and the low-Reynolds formulation, where turbulence equations are solved down to the wall, is applied for all turbulence models since the mesh provides $y^+$ values below unity.

\begin{figure}[h!]
\centering
\includegraphics[trim = 0 75 630 10 , clip, width = 0.55\textwidth]{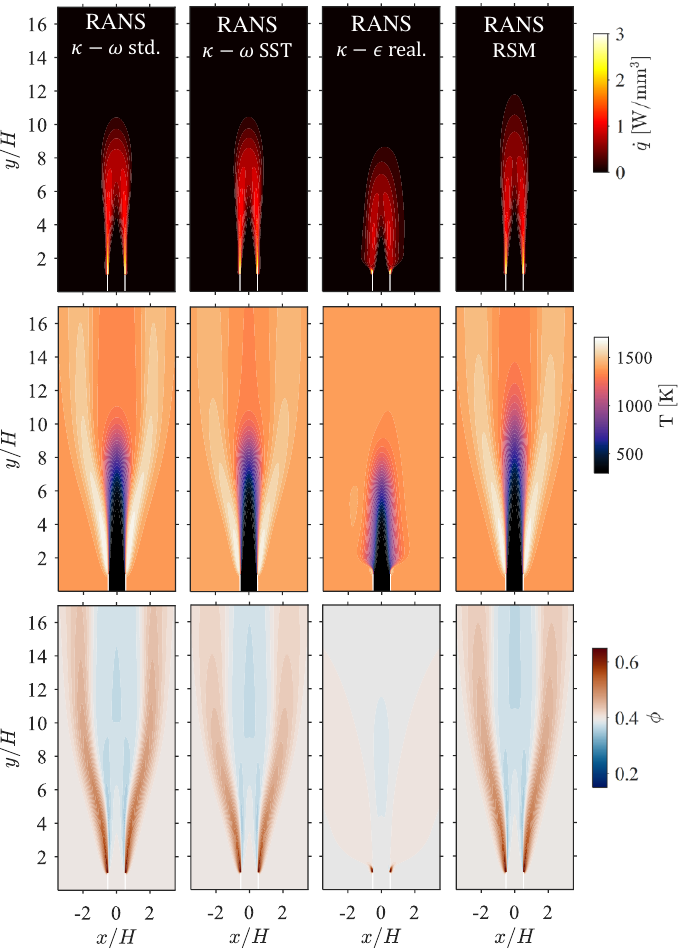}
\caption{Contours of heat release rate, temperature, and equivalence ratio for the hydrogen-air slot flame with Re = 5500 for alternative RANS turbulence models.}
\label{Fig_Re5500_TurbModel}
\end{figure}

The results of the simulations with the various turbulence models for the slot flame at Re = 5500 are presented in Figure~\ref{Fig_Re5500_TurbModel}. The largest difference is observed with the $\kappa$–$\epsilon$ model, which predicts a much shorter flame with almost no equivalence ratio variation and no super-adiabatic branches at the shear layer. A local enrichment can be seen at the flame base, but it is wiped out downstream. In contrast, the three remaining models show similar results.

\begin{figure*}[h!]
\centering
\includegraphics[trim = 0 5 325 0 ,clip,width=1.1\textwidth]{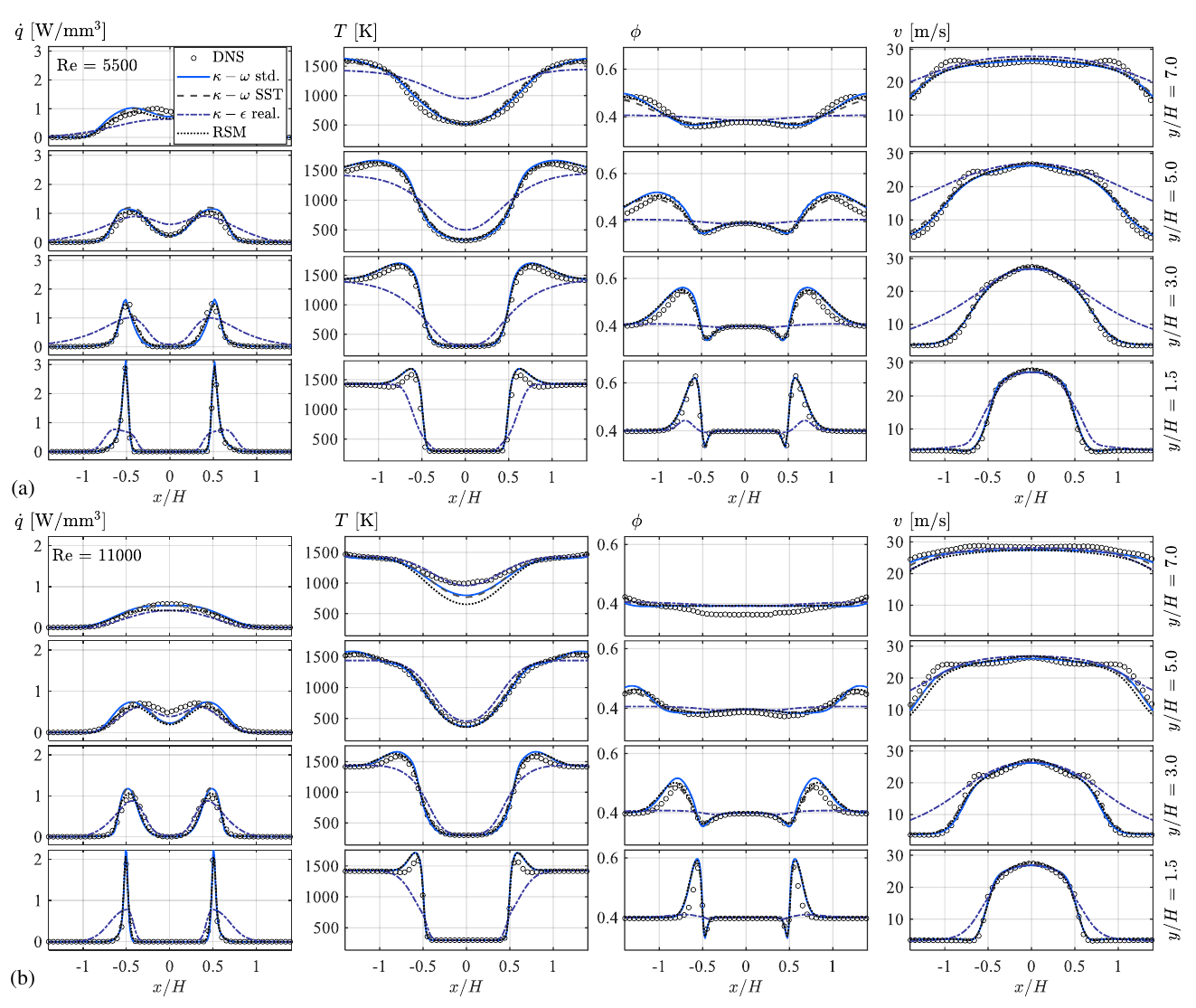}
\caption{Profiles of heat release rate, temperature, equivalence ratio, and $y$-velocity for the hydrogen-air slot flame with (a) Re = 5500 and (b) Re = 11000 at $y/H$ = 1.5, 3.0, 5.0, and 7.0. Favre-averaged DNS and RANS results obtained with different turbulence models.}
\label{Fig_profiles_TurbModel}
\end{figure*}

\begin{figure}[h!]
\centering
\includegraphics[trim = 0 330 623 5 , clip, width = 0.55\textwidth]{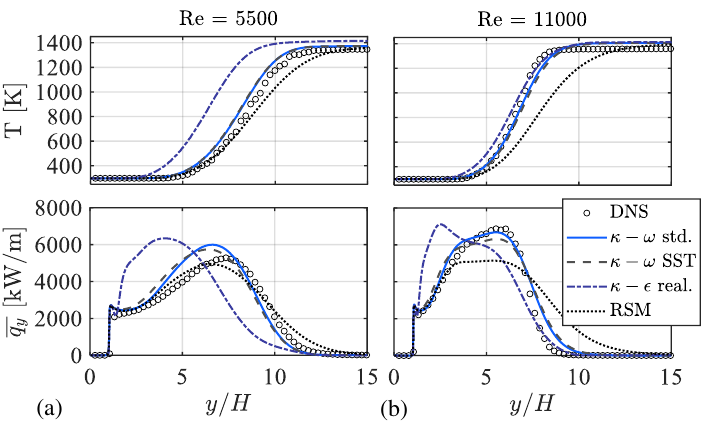}
\caption{Profiles of temperature ($x/H$ = 0.0) and streamwise distribution of the integrated heat release rate for the hydrogen-air slot flame with (a) Re = 5500 and (b) Re = 11000. Favre-averaged DNS and RANS with different turbulence models.}
\label{Fig_profiles_axis_TurbModel}
\end{figure}

A quantitative comparison with the DNS data as a reference is presented in Figures~\ref{Fig_profiles_TurbModel} and \ref{Fig_profiles_axis_TurbModel}. While the Reynolds Stress Model shows good agreement with the DNS in the shear layer, it predicts a longer flame. Although the RSM solves a transport equation for each Reynolds-stress component and is expected to better capture the shear-layer anisotropy of the slot jet, the values of the model constants $Sc_t$ and $C_{d}$ were selected based on the results obtained with the standard $\kappa$–$\omega$ model. Therefore, among the models evaluated, both $\kappa$–$\omega$ models are in better agreement with the DNS data.

The shorter flame length and smoother temperature and equivalence ratio fields obtained with the realizable $\kappa$–$\epsilon$ model indicate an overprediction of turbulent viscosity, which drives excessive turbulent mixing that smooths gradients and suppresses the non-unity Lewis number effects on local equivalence ratio variations. This is linked to a strong overshoot of the turbulent kinetic energy at the plane just downstream of the slot exit, as shown in ~\ref{A:1}. Turbulence--chemistry interaction is also affected by the inaccurate turbulent fields through the progress variable variance. The impact of turbulence on the source, diffusion, and scalar dissipation terms ultimately increases the variance experienced by the flame and, therefore, the local burning rate.

\section{Conclusions}

This work assessed a tabulated-chemistry model including preferential and differential diffusion (TC PD) within a RANS framework for a lean H$_2$–air slot jet at two Reynolds numbers. Using DNS as the reference, it was shown that the TC PD model captures both qualitatively and quantitatively the main characteristics of the turbulent hydrogen-lean flame, including the overall flame length and the regions of equivalence ratio variation and super-adiabatic temperature resulting from preferential and differential diffusion.

The effect of preferential diffusion in the TC model was assessed by comparing two additional versions of the TC model: one based on the unity Lewis number assumption (TC Le1) and another including preferential and differential diffusion only at the thermochemical level (TC PD-F). In addition to the turbulent slot jet flame, a one-dimensional freely propagating laminar flame was considered to evaluate the accuracy of the model implementation in the absence of turbulence and to help interpret the RANS results for the turbulent case. In the 1D laminar flame, only the TC PD model provided accurate predictions of flame speed, flame thickness, and flame structure, highlighting the importance of accounting for preferential and differential diffusion at both the thermochemical and transport levels. In the turbulent slot jet flame, although both the TC PD and TC PD-F models predicted flame lengths close to the DNS, including preferential and differential diffusion at the transport level (TC PD) is essential to capture all key features of lean premixed hydrogen flames.

Although the TC PD model performs well within a RANS framework, its accuracy is limited by the turbulence models used to represent unresolved transport, particularly the closure for the Reynolds stress tensor in the Favre-averaged Navier–Stokes equations, as demonstrated by comparisons with different turbulence models.

A sensitivity analysis of the TC PD model constants was also performed. Overall, turbulent Schmidt numbers in the range 0.4–0.6 and scalar dissipation constants in the range 2–3 provided good results. However, the optimum values of these constants are linked to the accuracy of the turbulence model. For example, underprediction of turbulent kinetic energy, and consequently turbulent viscosity, could be compensated by lowering the turbulent Schmidt number. Such error cancellation, however, cannot be generalized and is therefore not recommended. A systematic calibration approach could instead be pursued, where the turbulence model, turbulent Schmidt numbers, and scalar dissipation constants are calibrated independently and hierarchically using various reference cases, including cold-flow, passive-mixing, and reactive premixed and non-premixed conditions. Nevertheless, such calibration lies beyond the scope of the present work and is left for future studies.

Finally, heat losses were not considered in the present evaluation; however, their inclusion in the tabulated chemistry model with preferential and differential diffusion represents the next step toward applying the model in RANS simulations of practical combustion systems.

\section*{Acknowledgment}

This work has received funding from the European Union’s Horizon Europe research and innovation programme under grant agreement No. 101091456 (HyInHeat). EMF acknowledges the predoctoral grant Joan Oró-FI (2023 FI-1 00680) funded by AGAUR from the Secretariat of Universities and Research of the Department of Research and Universities of the Generalitat de Catalunya and the ESF+. DM acknowledges the Grant RYC2021-034654 funded by MICIU/AEI/10.13039/501100011033 and by ‘‘European Union NextGenerationEU/PRTR’’. EJPS acknowledges his AI4S fellowship within the ‘‘Generación D" initiative by Red.es, Ministerio para la Transformación Digital y de la Función Pública, for talent attraction (C005/24-ED CV1), funded by NextGenerationEU through PRTR.

\appendix

\section{Turbulence model evaluation in cold-flow}
\label{A:1}

\begin{figure}[h!]
\centering
\includegraphics[trim = 0 270 550 0  ,clip,width=0.55\textwidth]{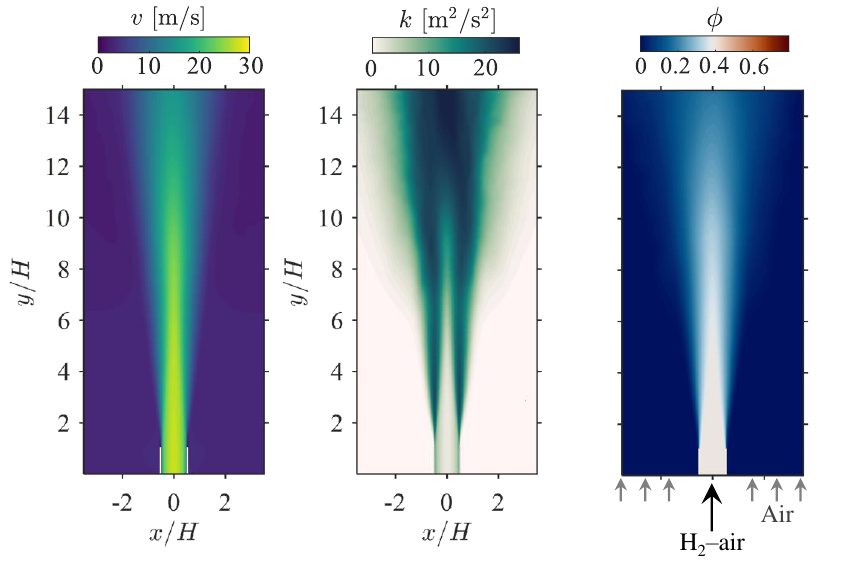}
\caption{Favre-average field of $y$-velocity, turbulent kinetic energy, and equivalence ratio from the DNS of the hydrogen-air slot cold-flow with Re = 5500.}
\label{Fig_DNS_coldFlow}
\end{figure}

To assess the predictions of the various turbulence models in the slot jet configuration,  RANS simulations of the cold-flow operation are performed. DNS data of the cold-flow operation are available for the case Re = 5500. Figure~\ref{Fig_DNS_coldFlow} shows the Favre-averaged fields from the DNS. As in the reacting case, the fully premixed H$_2$–air mixture is injected through the two parallel plates, while the coflow consists only of cold air, with the same velocity boundary conditions as in the reacting case. Although combustion does not occur, the transport of the mixture fraction still influences the results. Therefore, to remain consistent with the TC model setup, the model constants $Sc_t$ and $C_{d,z}$ are set to 0.5 and 3.0, respectively.

\begin{figure*}[h!]
\centering
\includegraphics[trim = 0 5 325 0  ,clip,width=1.1\textwidth]{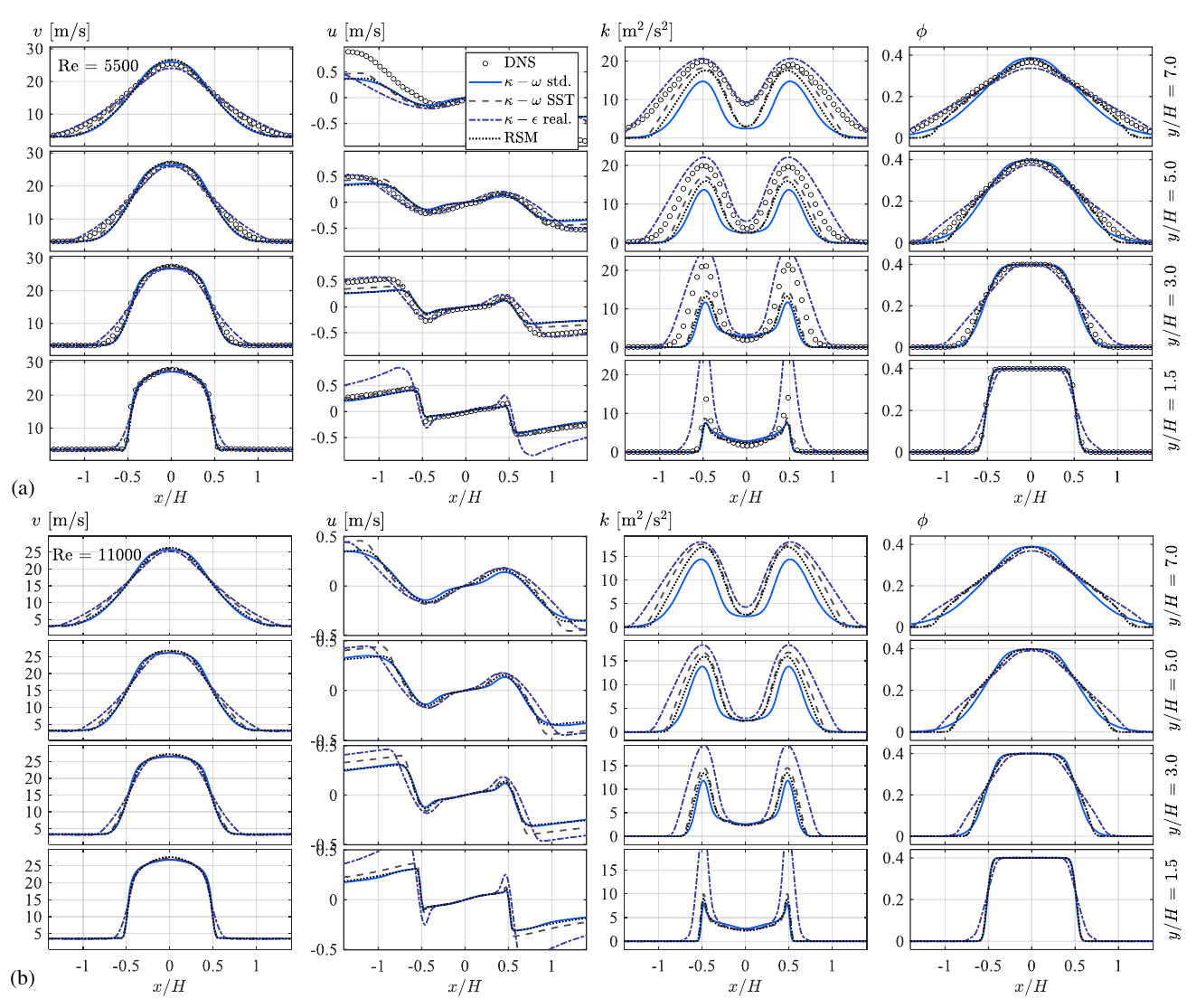}
\caption{Profiles of $y$-velocity, $x$-velocity, turbulent kinetic energy, and equivalence ratio for the hydrogen-air slot cold-flow with (a) Re = 5500 and (b) Re = 11000 at $y/H$ = 1.5, 3.0, 5.0, and 7.0.}
\label{Fig_DNSvRANS_profiles_coldFlow}
\end{figure*}

The results with the various turbulence models compared to the DNS data are presented in Figure~\ref{Fig_DNSvRANS_profiles_coldFlow}. The streamwise and cross-stream velocity components describe the mean jet shape and entrainment. The two $\kappa$–$\omega$ models and the RSM closely follow the DNS at all sections. The $\kappa$–$\epsilon$ model, however, while still not far from the DNS, shows stronger entrainment (marked by more negative $u$ lobes) and a slightly faster centerline decay.

A clear difference between the models can be seen for the turbulent kinetic energy. On the one hand, the $\kappa$–$\epsilon$ model strongly overshoots the turbulent kinetic energy at the plane just downstream of the slot exit, where turbulence originates almost entirely from the two plate boundary layers. Although the agreement with the DNS improves further downstream, the excessively high value of $\kappa$ at the flame base is sufficient to completely distort the flame shape and preferential diffusion effects in the reacting case, since equivalence ratio variation and super-adiabatic temperatures are mainly generated at the flame base and then convected downstream. This effect can also be seen in the transport of the equivalence ratio. As a passive scalar, $\phi$ spreads by molecular and turbulent diffusion while being convected downstream. Due to the overpredicted turbulent kinetic energy, the $\kappa$–$\epsilon$ model produces overly broadened (over-mixed) $\phi$ profiles compared to the DNS.

On the other hand, the $\kappa$–$\omega$ models and the RSM underpredict the turbulent kinetic energy. Nevertheless, the discrepancy is not as large as for the $\kappa$–$\epsilon$ model, and it allows steeper shear-layer gradients to be preserved while matching the DNS profile magnitude and shape much more closely. However, there is still room for improvement in the turbulence models to better predict turbulent kinetic energy and flow fields.

\bibliographystyle{model1-num-names}
\bibliography{TC_PD_RANS.bib}


\end{document}